\documentclass[prc,twocolumn,superscriptaddress,superscriptfootnote]{revtex4}
\usepackage{amssymb}

\usepackage{amsmath,amsfonts,amssymb,epsfig,color}

\begin{document}

\title{Simultaneous description of low-lying positive and negative parity bands \\
in heavy even-even nuclei}
\author{H. G. Ganev}
\affiliation{Joint Institute for Nuclear Research, 141980 Dubna,
Russia}

\setcounter{MaxMatrixCols}{10}

\begin{abstract}
The low-lying spectra including the first few excited positive and
negative parity bands of some heavy even-even nuclei from the rare
earth and actinide mass regions are investigated within the
framework of the symplectic Interacting Vector Boson Model with
Sp(12,$R$) dynamical symmetry group. Symplectic dynamical symmetries
allow the change of the number of excitation quanta or phonons
building the collective states providing for larger representation
spaces and richer subalgebraic structures to incorporate more
complex nuclear spectra. The theoretical predictions for the energy
levels and the electromagnetic transitions between the collective
states of the ground state band and $K^{\pi}=0^{-}$ band are
compared with experiment and some other collective models
incorporating octupole and/or dipole degrees of freedom. The energy
staggering which is a sensitive indicator of the octupole
correlations in the even-even nuclei is also calculated and compared
with experiment. The results obtained for the energy levels, energy
staggering and transition strengths reveal the relevance of the used
dynamical symmetry of the model for the simultaneous description of
both positive and negative parity low-lying collective bands.
\end{abstract}
\maketitle

PACS number(s): {21.10.Re, 21.60.Fw, 23.20.Lv, 27.60.+j}

\section{Introduction}

It is well known \cite{exp},\cite{BN} that in some mass regions
several bands of negative parity are observed in the low-lying
nuclear spectra in even-even nuclei, like $K^{\pi}=0^{-}$, $1^{-}$
and $2^{-}$ bands. The most well-studied of them is the
$K^{\pi}=0^{-}$ band, usually interpreted as an octupole vibrational
band, connected to the ground state band (GSB) by enhanced $E1$
transitions.

Negative parity states have been described within different
approaches mainly by inclusion of octupole or/and dipole degrees of
freedom. The bands of negative parity states are often associated
with the reflection asymmetry in the intrinsic frame of reference.
In the geometrical approach this is achieved  by including of the
$\alpha_{30} \equiv \beta_{3}$ deformation \cite{BM}. In the
Interacting Boson Model (IBM) \cite{IBM} the description of negative
states requires the introduction of $f$ or/and $p$ bosons with
negative parity in addition to the standard $s$ and $d$ bosons
($spdf$-IBM) \cite{spdfIBM1},\cite{spdfIBM2}. An alternative
interpretation of the low-lying negative parity states has been
provided in different cluster models \cite{CM1}-\cite{CM2} in which
the dipole degrees of freedom are related with the relative motion
of the clusters. Based on the Bohr Hamiltonian different critical
point symmetries (CPS) including axial quadrupole and octupole
deformations have been proposed \cite{Bizzeti1}-\cite{Bizzeti3},
\cite{Bonatsos} extending the concept of CPS introduced for the
description of positive parity states.

In this paper we present an algebraic approach, complementary to the
spdf-IBM \cite{spdfIBM1}, for the unified description of the
low-lying positive and negative parity bands in some even-even
nuclei from the rare earth and actinide mass regions within the
framework of the symplectic Interacting Vector Boson Model (IVBM)
with Sp(12,$R$) dynamical symmetry group \cite{IVBM}. The present
work is an extension of the approach proposed in Ref.\cite{GGG} for
the description of the ground state band and the "octupole"
($K^{\pi}=0^{-}$) band, often treated as a single ground state
alternating parity band. In this way we investigate simultaneously
the first few low-lying negative parity bands ($K^{\pi}= 0^{-}$,
$1^{-}$ and $2^{-}$) together with the first few positive parity
(ground state, $\beta$ and $\gamma$) bands. It is shown that the
negative parity bands arise along with the positive bands without
the introduction of any additional collective degrees of freedom.
Additionally, we calculate the strengths of the intraband $E2$
transitions in both the GSB and $K^{\pi}= 0^{-}$ band, as well as
the interband $E1$ transitions connecting the states of these two
bands. The energy staggering of the ground state alternating band
which is a sensitive indicator of the octupole correlations in the
even-even nuclei is also calculated and compared with experiment.

\section{Theoretical framework}
\subsection{The IVBM}

It was suggested by Bargmann and Moshinsky
\cite{BargMosh1},\cite{BargMosh2} that two types of bosons are
needed for the description of nuclear dynamics. It was shown there
that the consideration of only two-body system consisting of two
different interacting vector particles will suffice to give a
complete description of $N$ three-dimensional oscillators with a
quadrupole-quadrupole interaction. The latter can be considered as
the underlying basis in the algebraic construction of the
\emph{phenomenological} IVBM \cite{IVBM}.

The algebraic structure of the IVBM is realized in terms of creation
and annihilation operators of two kinds of vector bosons
$u_{m}^{\dag}(\alpha )$, $u_{m}(\alpha )$ ($m=0,\pm 1$), which
differ in an additional quantum number $\alpha=\pm1/2$ (or
$\alpha=p$ and $n$)$-$the projection of the $T-$spin (an analogue to
the $F-$spin of IBM-2 or the $I-$spin of the particle-hole IBM). All
bilinear combinations of the creation and annihilation operators of
the two vector bosons generate the boson representations of the
non-compact symplectic group $ Sp(12,R)$:
\begin{eqnarray}
F_{M}^{L}(\alpha,\beta) = {\sum}_{k,m}C_{1k1m}^{LM}u_{k}^{+}(\alpha)u_{m}^{+}(\beta), \label{Fs} \\
G_{M}^{L}(\alpha,\beta) ={\sum }_{k,m}C_{1k1m}^{LM}u_{k}(\alpha)u_{m}(\beta), \label{Gs} \\
A_{M}^{L}(\alpha,\beta)={\sum}_{k,m}C_{1k1m}^{LM}u_{k}^{+}(\alpha)u_{m}(\beta),
\label{numgen}
\end{eqnarray}
where $C_{1k1m}^{LM}$, which are the usual Clebsch-Gordan
coefficients for $L=0,1,2$ and $M=-L,-L+1,...L$, define the
transformation properties of (\ref{Fs}),(\ref{Gs}) and
(\ref{numgen}) under rotations. The number preserving operators
(\ref{numgen}) generate the U(6) group, while by adding the pair
creation (\ref{Fs}) and annihilation (\ref{Gs}) operators we
generate the non-compact Sp(12,$R$) which is the dynamical group of
the IVBM. Its irreducible representations are infinite dimensional.
We also introduce the following notations for the two bosons:
$u_{m}^{\dag}(\alpha=1/2)= p^{\dag}_{m}$ and
$u_{m}^{\dag}(\alpha=-1/2)=n^{\dag}_{m}$.

Symplectic dynamical symmetries allow the change of the number of
bosons, elementary excitations or phonons $N$, providing for richer
subalgebraic structures and larger representation spaces to
accommodate more structural effects. Dynamical symmetry group
Sp(12,$R$) contains both compact and non-compact substructures,
defined by different reduction chains.

\subsection{Dynamical symmetry}

We consider the following chain \cite{IVBM},\cite{GGG}
\begin{align}
&Sp(12,R) \supset U(6) \supset \ SU(3) \ \otimes \ U(2) \quad
\supset SO(3) \otimes U(1), \notag\\
&\quad\quad\quad\quad\quad\quad [N]_{6} \quad (\lambda,\mu)
\Longleftrightarrow (N,T) \ \ K \quad L \quad\quad\quad T_{0} \
\label{DS}
\end{align}
where below the different subgroups the quantum numbers
characterizing their irreducible representations are given. The
generators of different subgroups in Eq.(\ref{DS}) are expressed in
terms of the number-preserving operators (\ref{numgen}). The number
operator
\begin{equation}
N = p^{\dag} \cdot p + n^{\dag} \cdot n = N_{p} + N_{n} \label{Ntot}
\end{equation}
is the linear invariant of the U(6), as well as U(3) and U(2)
algebras. The SU(3) algebra is generated by the components of the
angular momentum
\begin{equation}
L_{M} = -\sqrt{2}\sum_{\alpha} A^{1}_{M}(\alpha,\alpha) \label{L}
\end{equation}
and Elliott's quadrupole
\begin{equation}
Q_{M} = \sqrt{6}\sum_{\alpha} A^{2}_{M}(\alpha,\alpha) \label{Q}
\end{equation}
operators. The $T$-spin operators:
\begin{align}
&T_{+1} = -\frac{1}{\sqrt{2}}p^{\dag} \cdot n, \label{Tp} \\
&T_{-1} = \frac{1}{\sqrt{2}}n^{\dag} \cdot p, \label{Tm} \\
&T_{0} = \frac{1}{2}(p^{\dag} \cdot p - n^{\dag} \cdot n) \label{T0}
\end{align}
together with the number operator (\ref{Ntot}) generate the U(2)
algebra.

Within the symmetric irreducible representation $[N]_{6}$ of U(6)
the groups SU(3) and U(2) are mutually complementary \cite{MQ}, i.e.
the quantum numbers $(\lambda,\mu)$ are related with $(N,T)$ in the
following way $N = \lambda +2\mu$ and $T = \lambda/2$. Making use of
the latter we can write the basis as
\begin{equation}
\mid \lbrack N]_{6};(\lambda ,\mu );K,L;T_{0}\rangle =\mid
(N,T);K,L;T_{0}\rangle  \label{basis}
\end{equation}
The ground state of the system is:
\begin{equation}
\mid 0\text{\ \ }\rangle =\mid (N=0,T=0);K=0,L=0;T_{0}=0\text{
}\rangle \label{GS}
\end{equation}
which is the vacuum state for the Sp(12,$R$) group.

The basis states associated with the even irreducible representation
of the Sp(12,$R$) can be constructed by the application of powers of
raising generators $F_{M}^{L}(\alpha ,\beta)$ of the same group on
the vacuum. Each raising operator will increase the number of bosons
$N$ by two. The resulting infinite set of basis states so obtained
is denoted as (\ref{basis}) and is shown in Table \ref{BS}. Each row
(fixed $N$) of the table corresponds to a given U(6) irrep, whereas
each cell represents the SU(3) irrep contained in the corresponding
U(6) one. For fixed $N$, the possible values for the $T$-spin are
$T=\frac{N}{2},\frac{N}{2}-1,...$ $0$ and are given in the column
next to the respective value of $N$. Thus when $N$ and $T$ are
fixed, $2T+1$ equivalent representations $(\lambda,\mu)$ of the
group SU(3) arise. Each of them is labeled by the eigenvalues of the
operator $T_{0}:-T,-T+1,...,T,$ defining the columns of Table
\ref{BS}. The values of the angular momentum contained in a certain
SU(3) representation $(\lambda,\mu)$ are obtained by means of
standard reduction rules for the chain SU(3) $\supset$ SO(3)
\cite{Elliott}:
\begin{align}
&K=\min (\lambda,\mu),\min (\lambda,\mu)-2,...,0~(1)  \notag \\
&L=\max (\lambda,\mu),\max (\lambda,\mu)-2,...,0~(1);K=0
\label{su3o3}
\\
&L=K,K+1,...,K+\max (\lambda ,\mu );K\neq 0.  \notag
\end{align}
The multiplicity index $K$ appearing in this reduction is related to
the projection of $L$ in the body fixed frame and is used with the
parity ($\pi$) to label the different bands ($K^{\pi}$) in the
energy spectra of the nuclei.

\begin{table}[h]
\caption{Classification of the basis states.} \label{BS}
\smallskip \centering%
\begin{tabular}{||l|l||llll|l|llll||}
\hline\hline
$N$ & $T$ & $T_{0}\cdots $ & \multicolumn{1}{|l}{$-3$} & \multicolumn{1}{|l}{%
$-2$} & \multicolumn{1}{|l|}{$-1$} & $\ \ \ 0$ & $\ \ 1$ &
\multicolumn{1}{|l}{$\ \ 2$} & \multicolumn{1}{|l}{$\ \ 3$} &
\multicolumn{1}{|l||}{$\cdots $} \\ \hline\hline $0$ & $0$ &  &  & &
& $(0,0)$ &  &  &  &  \\ \hline $2$ & $1$ &  &  &  &
\multicolumn{1}{|l|}{$(2,0)$} & $(2,0)$ & $(2,0)$ &
\multicolumn{1}{|l}{} &  &  \\ \cline{6-8} & $0$ &  &  &  &  &
$(0,1)$ &  &  &  &  \\ \hline & $2$ &  &  &
\multicolumn{1}{|l}{$(4,0)$} & \multicolumn{1}{|l|}{$(4,0)$} &
$(4,0)$ & $(4,0)$ & \multicolumn{1}{|l}{$(4,0)$} &
\multicolumn{1}{|l}{} &
\\ \cline{5-9}
$4$ & $1$ &  &  &  & \multicolumn{1}{|l|}{$(2,1)$} & $(2,1)$ &
$(2,1)$ & \multicolumn{1}{|l}{} &  &  \\ \cline{6-8} & $0$ &  &  & &
& $(0,2)$ &  &  &  &  \\ \hline & $3$ &  &
\multicolumn{1}{|l}{$(6,0)$} & \multicolumn{1}{|l}{$(6,0)$} &
\multicolumn{1}{|l|}{$(6,0)$} & $(6,0)$ & $(6,0)$ & \multicolumn{1}{|l}{$%
(6,0)$} & \multicolumn{1}{|l}{$(6,0)$} & \multicolumn{1}{|l||}{} \\
\cline{4-10}
$6$ & $2$ &  &  & \multicolumn{1}{|l}{$(4,1)$} & \multicolumn{1}{|l|}{$(4,1)$%
} & $(4,1)$ & $(4,1)$ & \multicolumn{1}{|l}{$(4,1)$} &
\multicolumn{1}{|l}{} &  \\ \cline{5-9} & $1$ &  &  &  &
\multicolumn{1}{|l|}{$(2,2)$} & $(2,2)$ & $(2,2)$ &
\multicolumn{1}{|l}{} &  &  \\ \cline{6-8} & $0$ &  &  &  &  &
$(0,3)$ &  &  &  &  \\ \hline $\cdots $ & $\cdots $ & $\cdots $ &
\multicolumn{1}{|l}{$\cdots $} & \multicolumn{1}{|l}{$\cdots $} &
\multicolumn{1}{|l|}{$\cdots $} & $\cdots $ & $\cdots $ &
\multicolumn{1}{|l}{$\cdots $} & \multicolumn{1}{|l}{$\cdots $}
& \multicolumn{1}{|l||}{$\cdots $}%
\end{tabular}
\end{table}

\subsection{The Hamiltonian}

We use the following Hamiltonian \cite{GGG}:
\begin{equation}
H= aN + bN^{2} +\alpha_{3}T^{2} +\beta_{3}L^{2}+\alpha_{1}T^{2}_{0}
\label{H}
\end{equation}
expressed in terms of the first and second order invariant operators
of the different subgroups in the chain (\ref{DS}). It is obviously
diagonal in the basis (\ref{basis})  and its eigenvalues are just
the energies of the nuclear system:
\begin{equation}
E(N,L,T,T_{0})= aN + bN^{2} +\alpha_{3}T(T+1)
+\beta_{3}L(L+1)+\alpha_{1}T^{2}_{0}. \label{En}
\end{equation}
The \ energy of the \ ground state (\ref{GS}) of the \ system is
obviously \ $0$.

\section{Application}

In our application, the most important point is the identification
of the experimentally observed states with a certain subset of the
basis states (\ref{basis}). In this regard, the following two points
are of importance. First, as we noted the irreducible
representations of Sp(12,$R$) are infinite dimensional. Then the
truncation of the model space to a finite-dimensional subspace of
physically meaningful basis states revealing the collective
properties of states described is required. It turns out that such
an appropriate set of states is given by the so called "stretched
states" \cite{str}, which represent dominant SU(3) multiplets in the
low-lying collective states \cite{str}. In the present application
we use the following type of stretched states defined as the SU(3)
states of the type $(\lambda,\mu )=(\lambda_{0},\mu_{0}+k)$, where
$k=0,2,4, \ldots$. In the symplectic IVBM the change of the number
$k$, which is related in the applications to the angular momentum
$L$ of the states, gives rise to the collective bands.

The second point concerns the parity of the state. We assume that
the one type of two vector bosons, say $p$-boson, transforms under
space reflections as a pseudovector, while the other - $n$-boson -
transforms as a vector. The latter assumes that the creation
operators of the two vector bosons $p^{\dag}_{m}$ and $n^{\dag}_{m}$
can be considered as acting separately in the two adjacent major
oscillator shells of opposite parity, creating in this way two
different elementary excitations ("Elliott quarks", see
\cite{Lipkin}) with opposite parity from which the collective states
are built out. So, we define the parity of the considered collective
state as  $\pi = (-1)^{N_{n}}$ which generalizes our previous
definition of the parity $\pi = (-1)^{T}$ given in Ref.\cite{GGG}.
This allows us to describe both positive and negative parity states
in the IVBM on the same footing without introducing of any
additional collective degrees of freedom.

In this way for example, the states of the ground state band are
mapped onto the SU(3) multiplets $(0,L)$ ($ T= 0$, $T_{0} = 0$) with
$L = 0, 2, 4, \ldots$, whereas those of $K^{\pi}=0^{-}$ band onto
the SU(3) multiplets $(2,L)$ ($T = 1$, $T_{0} = 1$) with $L = 1, 3,
5, \ldots$. The latter mapping slightly differs from that used in
Ref.\cite{GGG} with ($T = 1$, $T_{0} = 0$) because of the parity
definition. We note that although the set of used SU(3) states in
\cite{GGG} and in the present approach is identical, in order to
take proper into account the parity of the collective states, we
need appropriate values of both $T$ and $T_{0}$. Note that the SU(3)
degeneracy within a given U(6) irrep is lifted by its mutually
complimentary group U(2). The same type of the stretched states
$(\lambda,\mu )=(\lambda_{0},\mu_{0}+k)$ are also used for other
bands under consideration.

\subsection{The energy spectra}

\begin{widetext}

\begin{figure}[h]\centering
\includegraphics[width=78mm]{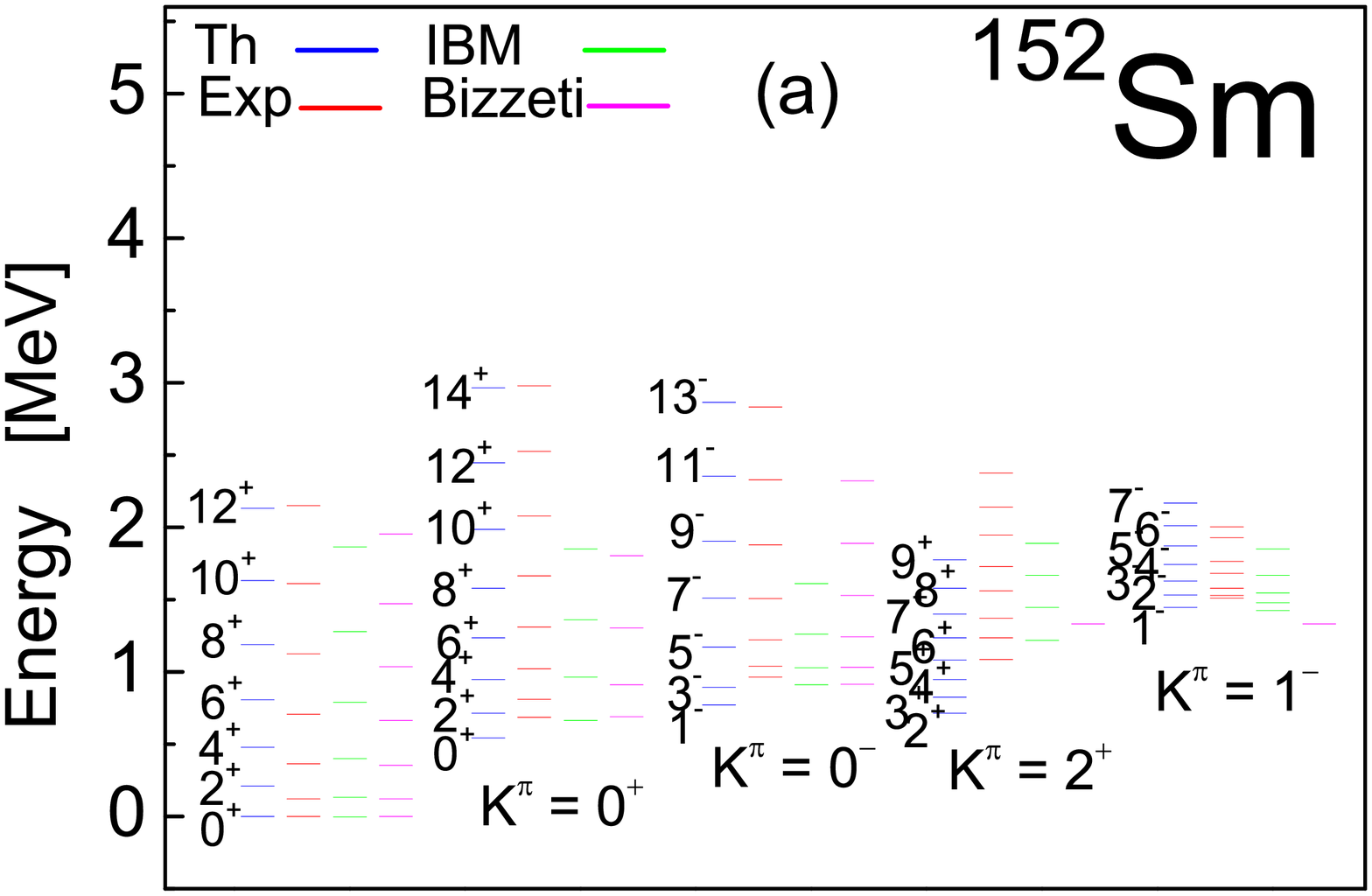}\hspace{1.mm}
\includegraphics[width=78mm]{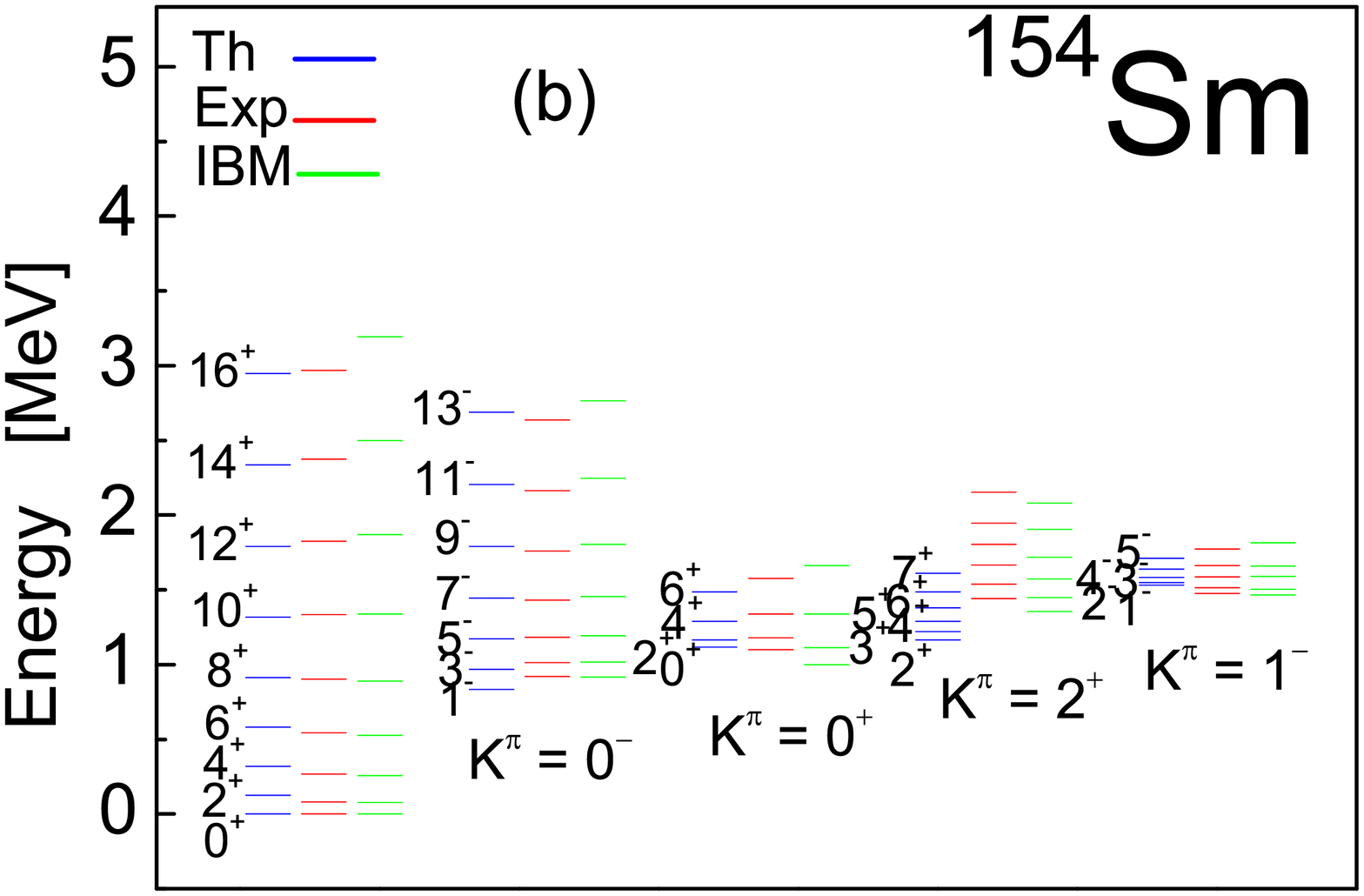}\hspace{1.mm}
\includegraphics[width=78mm]{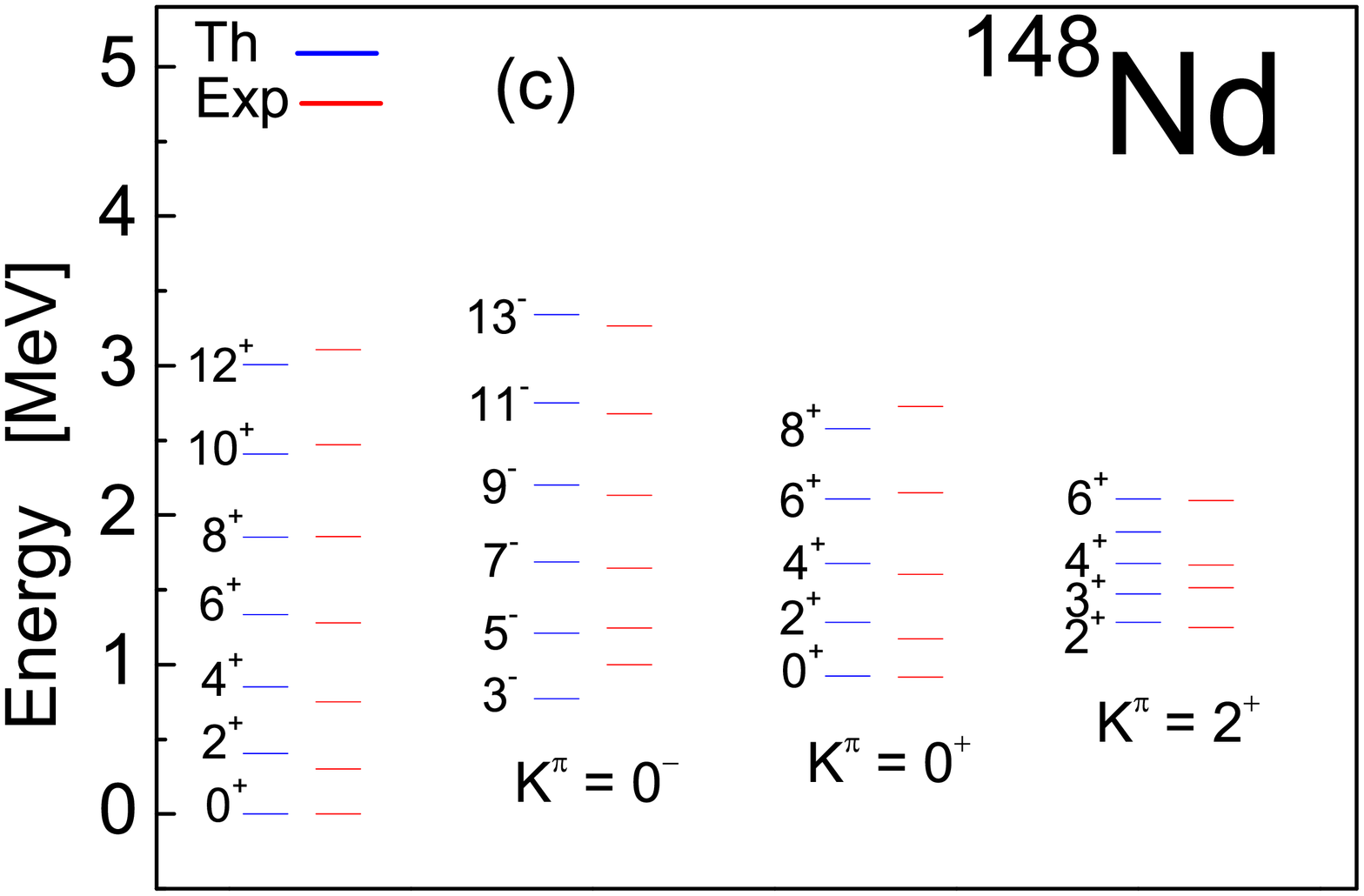}\hspace{1.mm}
\includegraphics[width=78mm]{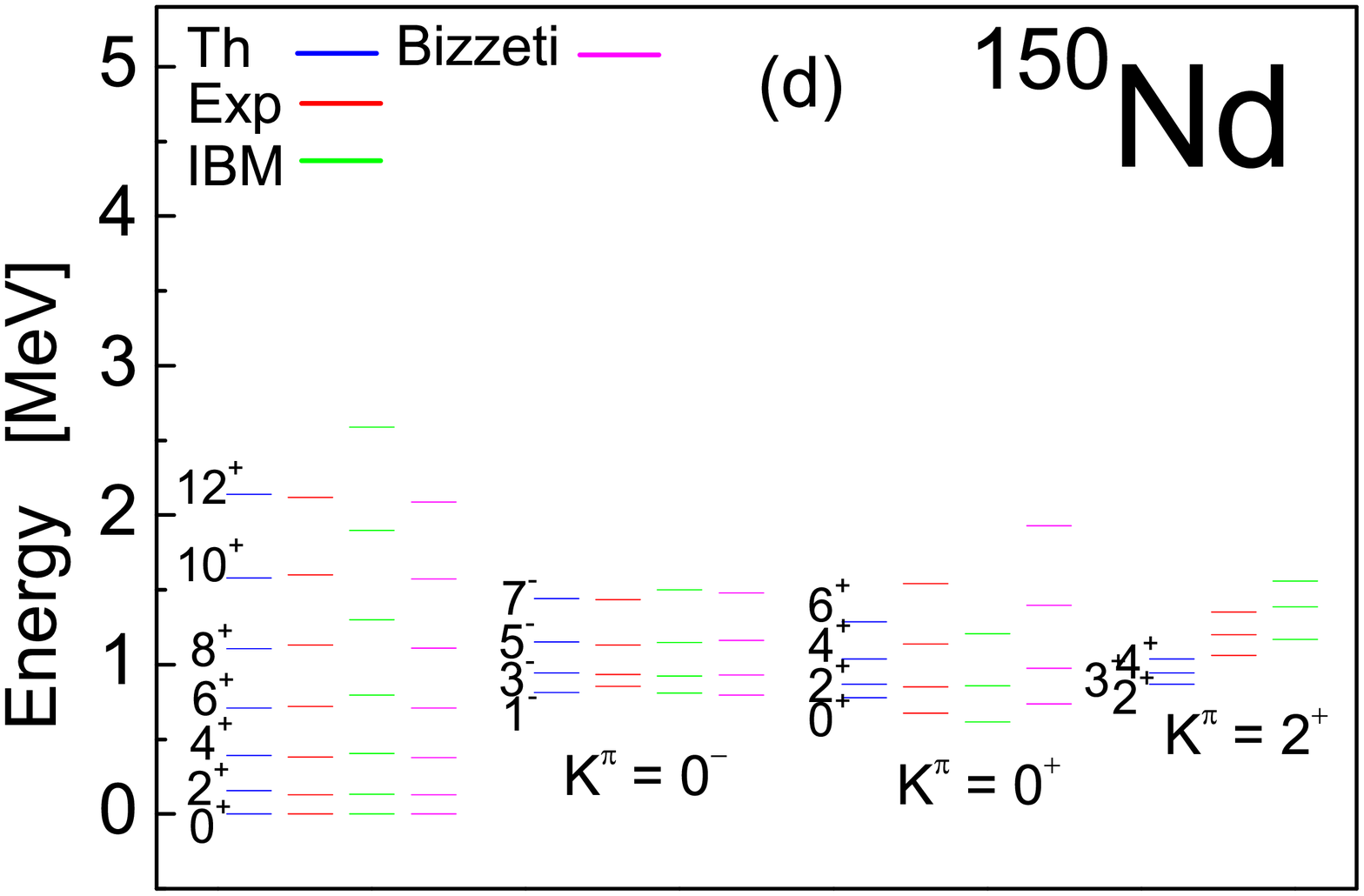}\hspace{1.mm}
\includegraphics[width=78mm]{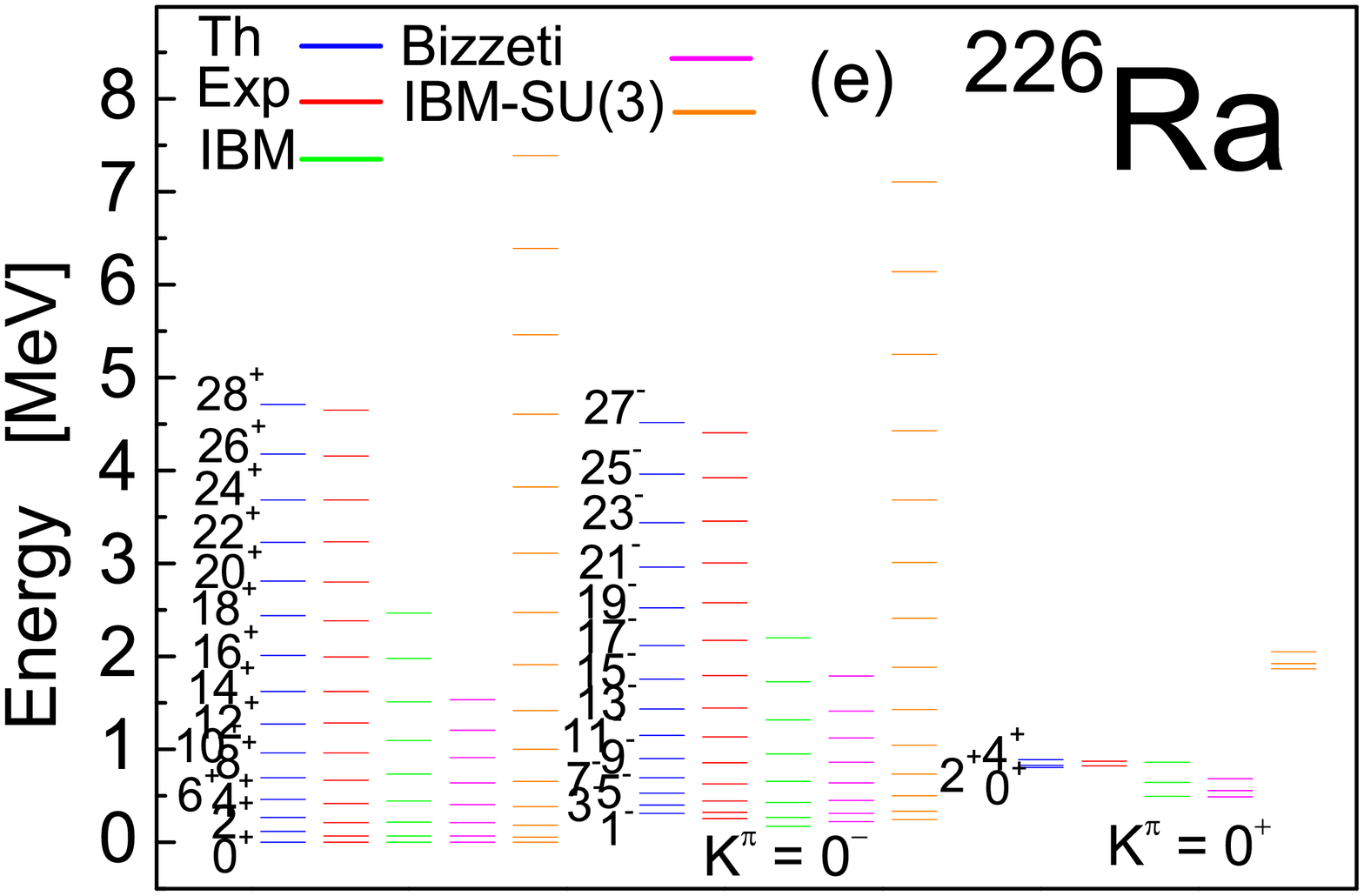}\hspace{1.mm}
\includegraphics[width=78mm]{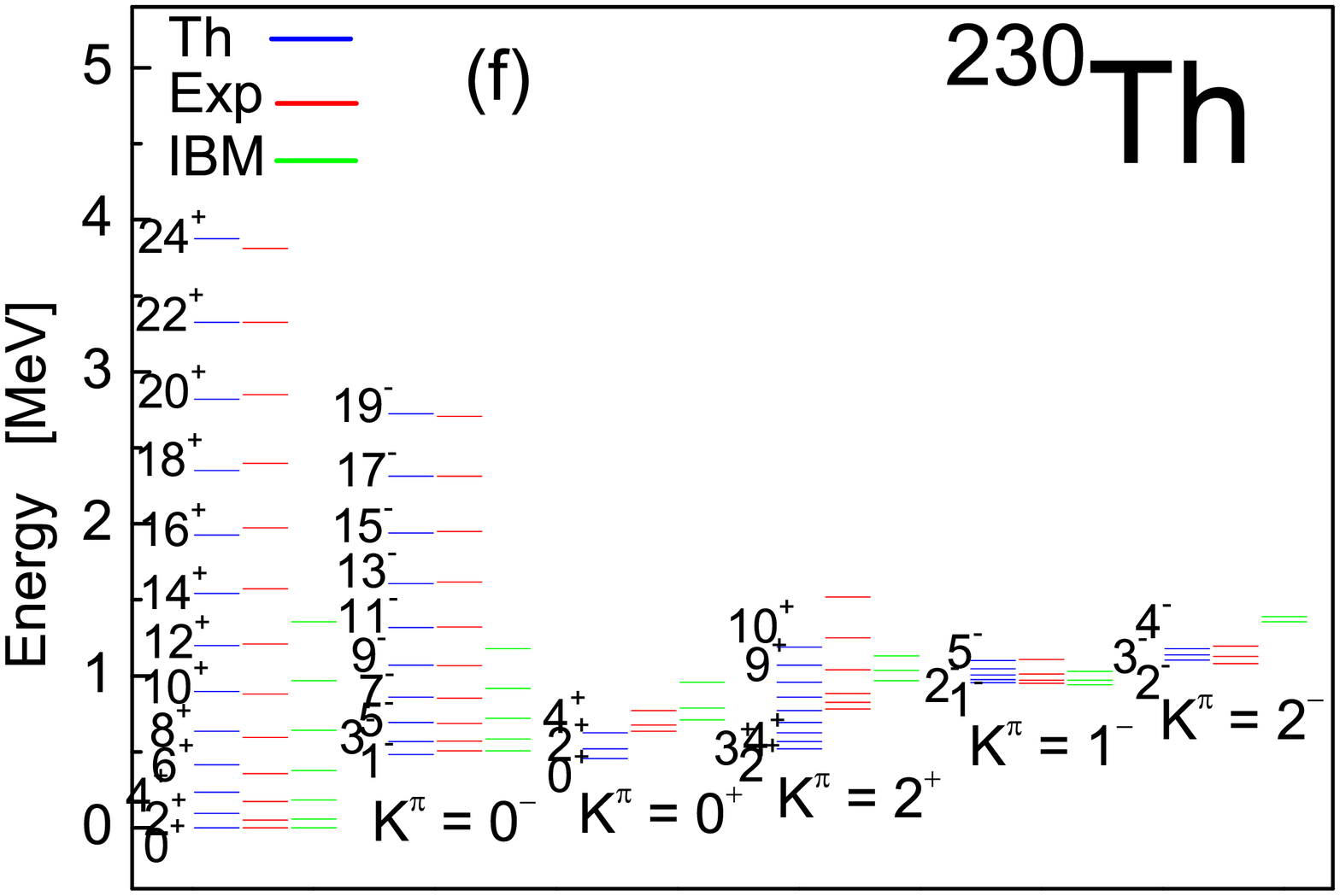}
\caption{(Color online) Comparison of the theoretical energies for
the low-lying positive and negative parity bands in $^{152}$Sm,
$^{154}$Sm, $^{148}$Nd, $^{150}$Nd, $^{226}$Ra and $^{230}$Th with
experiment and some other collective models incorporating octupole
or/and dipole degrees of freedom.} \label{Energies}
\end{figure}

\end{widetext}

We consider the first few excited low-lying positive (ground state,
$\beta$, $\gamma$) and negative ($K^{\pi} = 0^{-}$, $1^{-}$,
$2^{-}$) parity bands of some nuclei from the rare earth and light
actinide regions for which there is enough experimental data on $E1$
and $E2$ transitions.

In Fig.\ref{Energies} we compare our theoretical predictions for the
energies of the first excited positive and negative parity bands
observed in $^{152}$Sm, $^{154}$Sm, $^{148}$Nd, $^{150}$Nd,
$^{226}$Ra and $^{230}$Th with experiment \cite{exp} and the results
obtained by the diagonalization of the $spdf$-IBM Hamiltonian
\cite{Babilon} ($^{152}$Sm, $^{154}$Sm), \cite{spdfIBM-Nd150}
($^{150}$Nd), \cite{spdfIBM-Ra226} ($^{226}$Ra, $^{230}$Th). For the
$^{152}$Sm, $^{150}$Nd and $^{226}$Ra isotopes, the predictions of
the CPS approach \cite{Bizzeti3},\cite{Bizzeti2} in which the
octupole degrees of freedom are included together with the
quadrupole ones are also shown. In the case of $^{226}$Ra the
results of the pure SU(3) dynamical limit of the $spdf$-IBM are
shown as well. The calculations in the SU(3) limit of $spdf$-IBM are
performed using the Hamiltonian and matrix elements given in
\cite{SU3-spdfIBM}. The values of the model parameters obtained in
the fitting procedure are given in Table \ref{pars}.

\begin{table}[h]
\caption{The values of the model parameters (in MeV).} \label{pars}
\smallskip \centering%
\begin{tabular}{||l||l|l|l|l|l||}
\hline\hline
$Nucleus$ & $a$ & $b$ & $\alpha _{3}$ & $\beta _{3}$ & $\alpha _{1}$ \\
\hline\hline
$^{152}$Sm & $0.02792$ & $-0.00176$ & $0.10948$ & $0.01551$ & $0.46287$ \\
\hline
$^{154}$Sm & $0.01476$ & $-0.00153$ & $0.06864$ & $0.01486$ & $0.63245$ \\
\hline
$^{148}$Nd & $0.09149$ & $-0.00155$ & $0.09725$ & $0.01094$ & $-0.18550$ \\
\hline
$^{150}$Nd & $0.01572$ & $-0.00413$ & $0.95750$ & $0.02656$ & $-1.1522$ \\
\hline
$^{226}$Ra & $0.01581$ & $-0.00278$ & $0.12640$ & $0.01600$ & $0.00523$ \\
\hline
$^{230}$Th & $0.01248$ & $-0.00204$ & $0.15437$ & $0.01331$ & $0.13035$ \\
\hline\hline
\end{tabular}%
\end{table}

The $^{152}$Sm and $^{150}$Nd isotopes in the positive parity part
(GSB and $\beta$ band) of the spectrum are considered as examples of
the X(5) critical point symmetry \cite{CPSsm152}. The nucleus
$^{226}$Ra is considered in the literature as possessing stable
octupole shape. The $^{230}$Th is considered as an octupole soft
nucleus in a recent constrained self-consistent relativistic
mean-field calculations \cite{Nomura2013}.

One sees that the IVBM describes reasonably well the structure of
low-lying excited states of the first few bands of positive and
negative parity up to high angular momenta for the all nuclei under
consideration. Note that in the case of $^{226}$Ra, the experimental
data show large deviations from the rotational $L(L+1)$ rule (SU(3)
limit of the spdf-IBM) for both the ground state and $K^{\pi} =
0^{-}$ bands despite the fact that $R_{4/2} = 3.13$.

\begin{widetext}

\begin{figure}[h]\centering
\includegraphics[width=70mm]{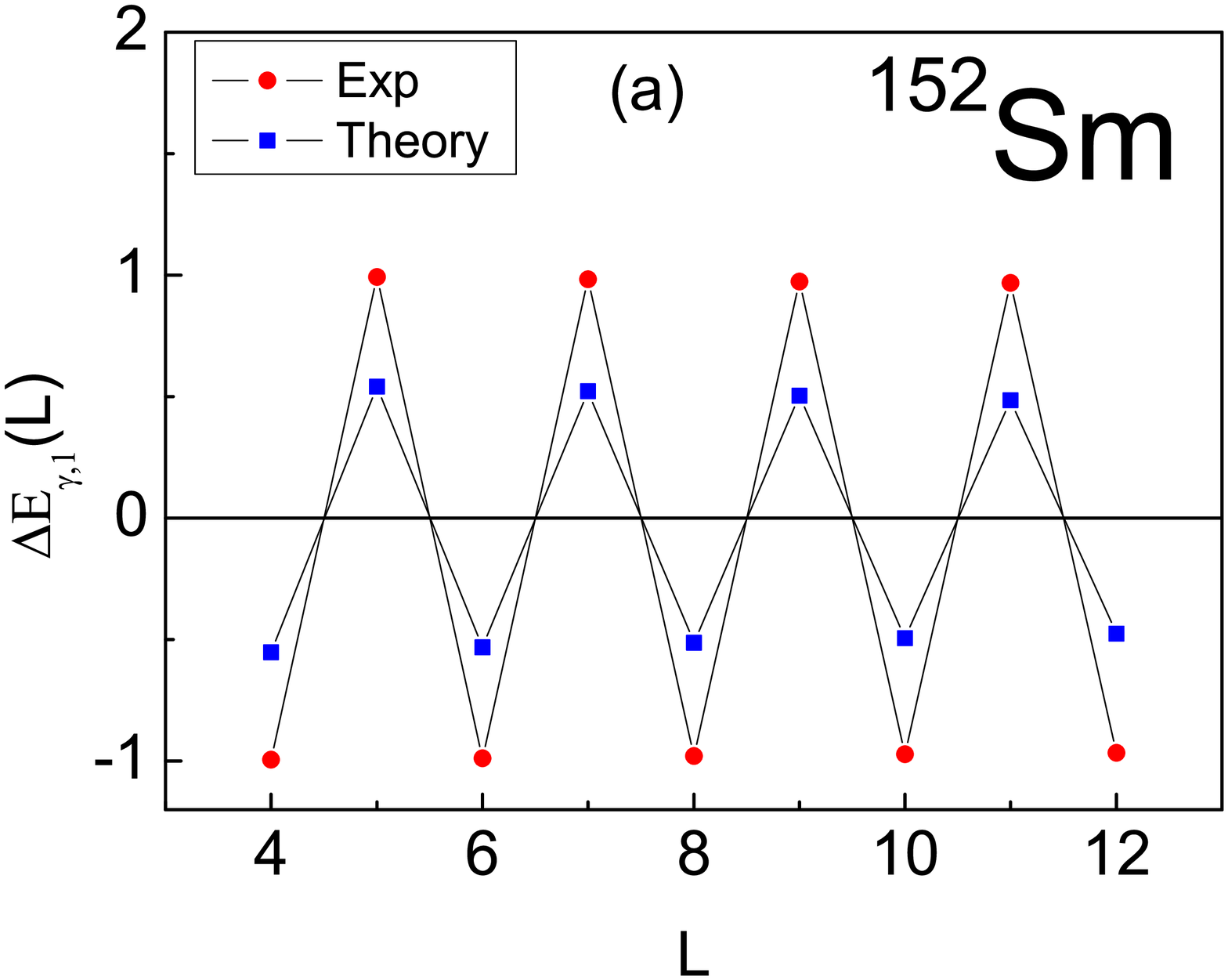}\hspace{1.mm}
\includegraphics[width=70mm]{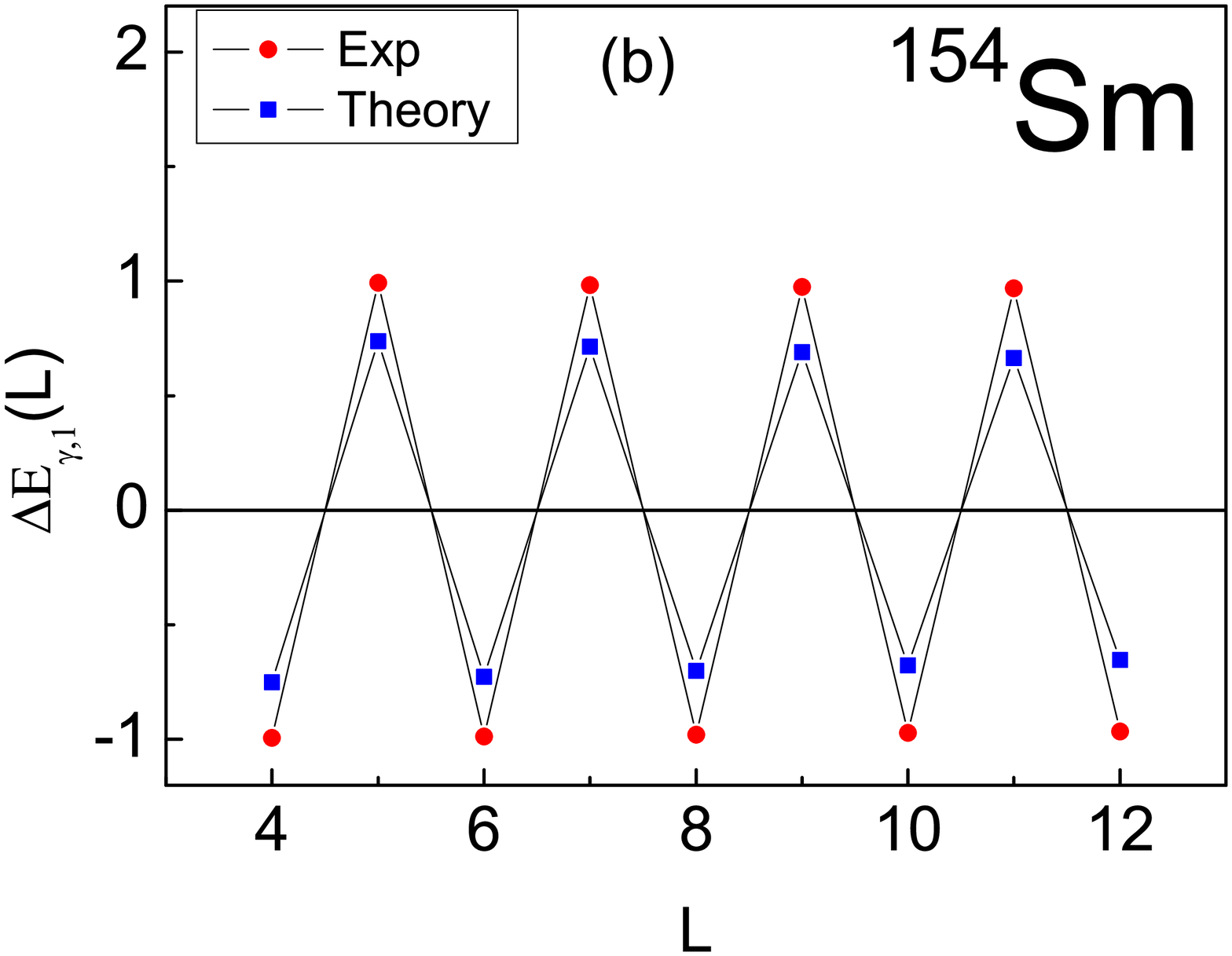}\hspace{1.mm}
\includegraphics[width=70mm]{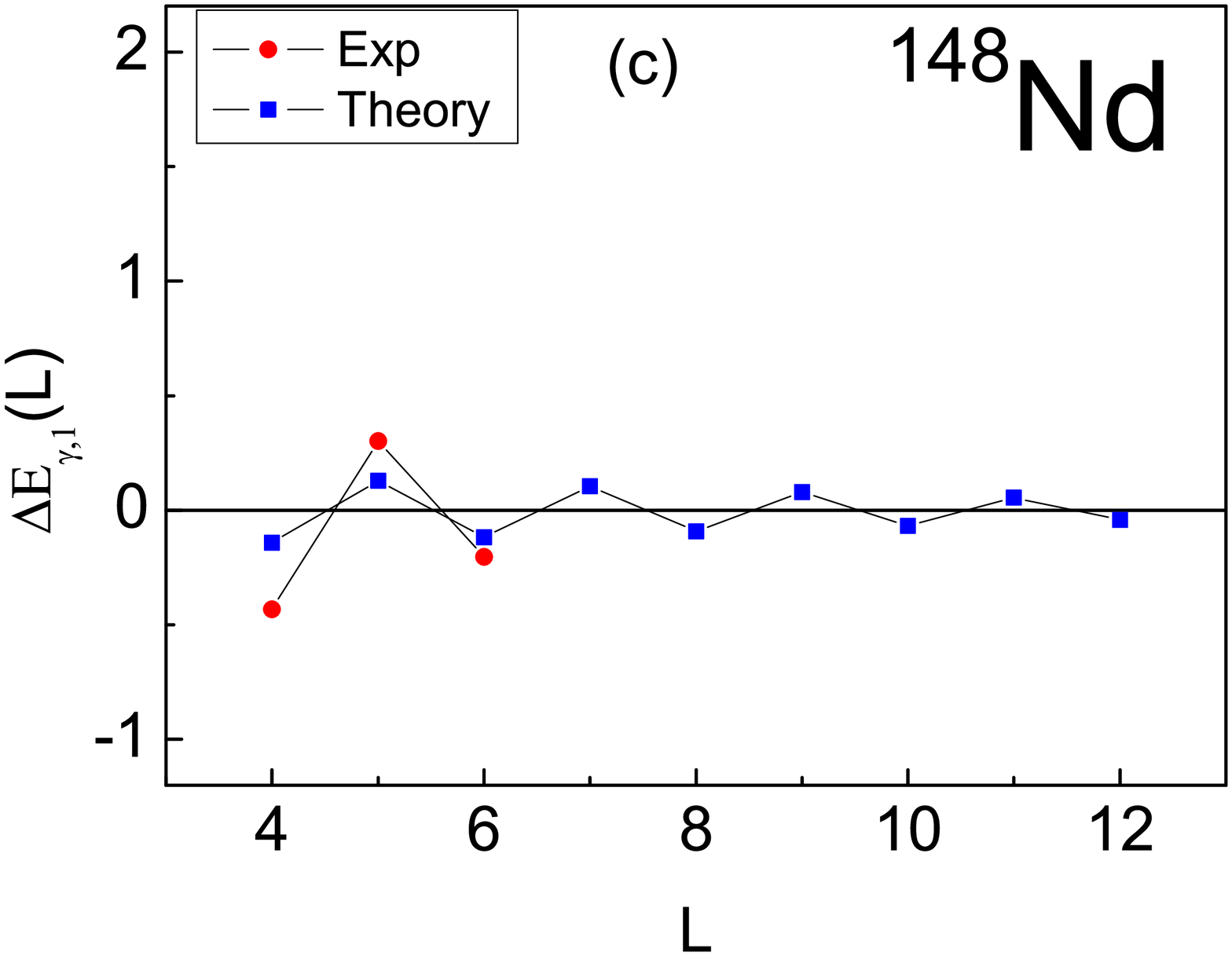}\hspace{1.mm}
\includegraphics[width=70mm]{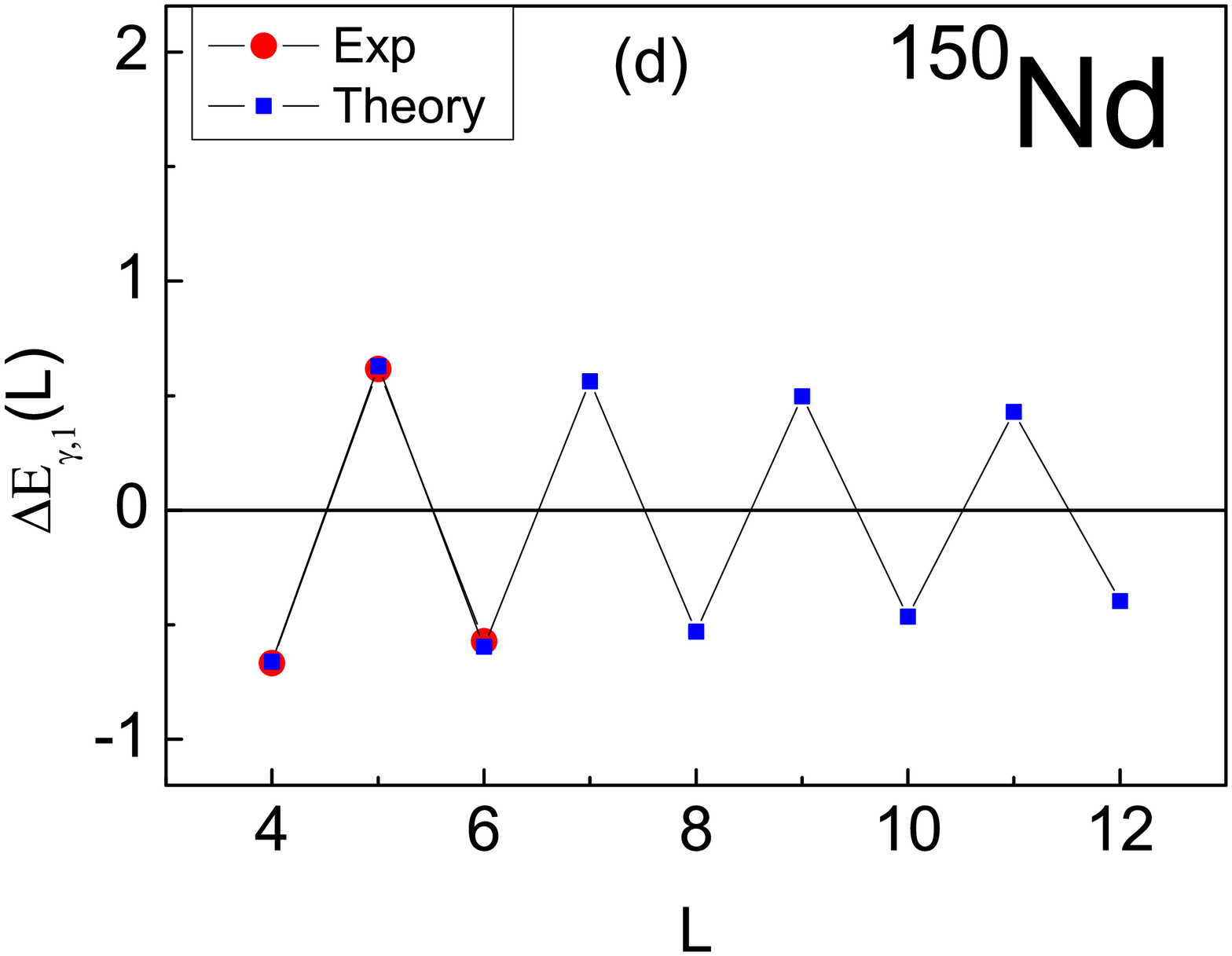}\hspace{1.mm}
\includegraphics[width=70mm]{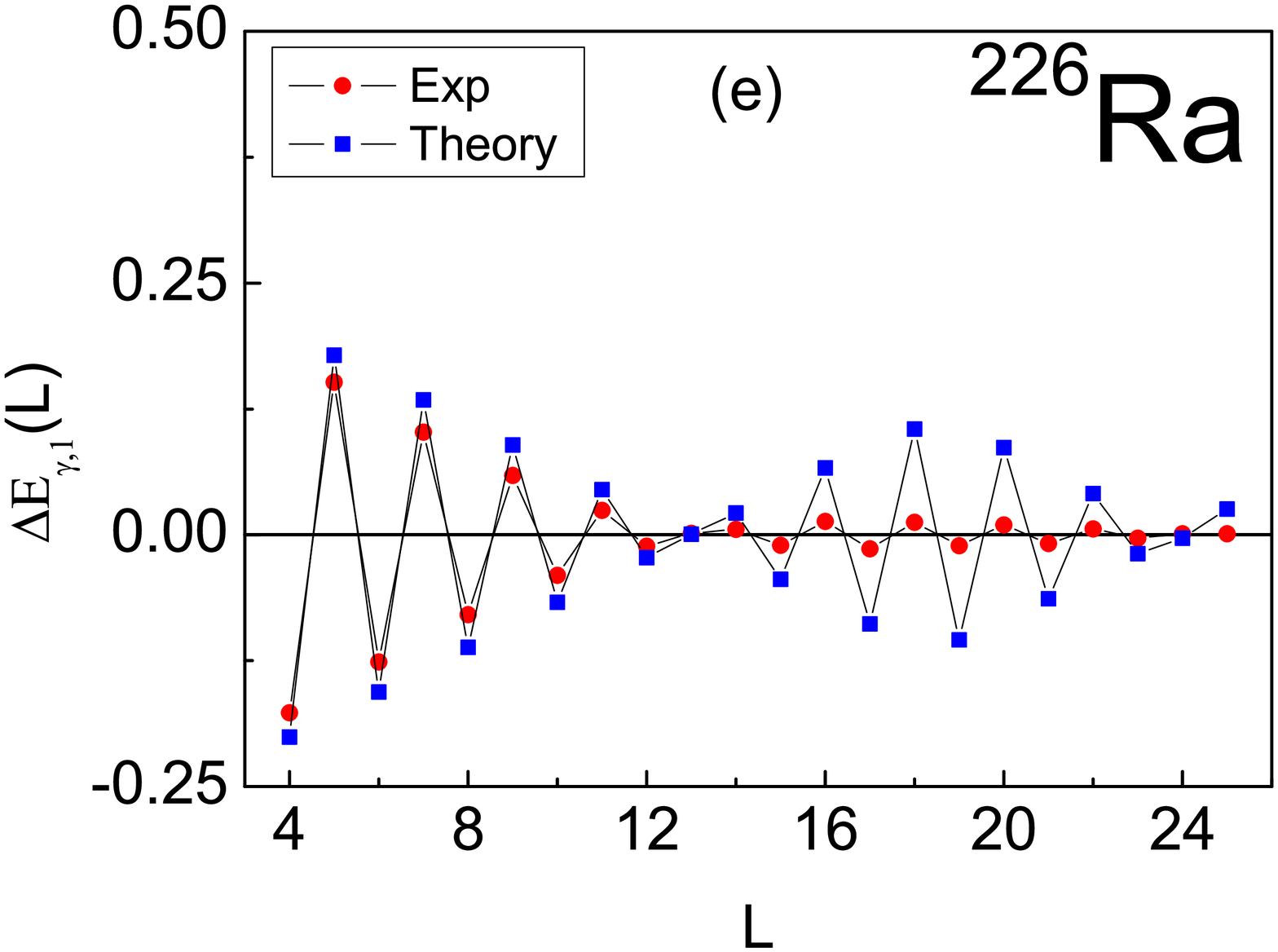}\hspace{1.mm}
\includegraphics[width=70mm]{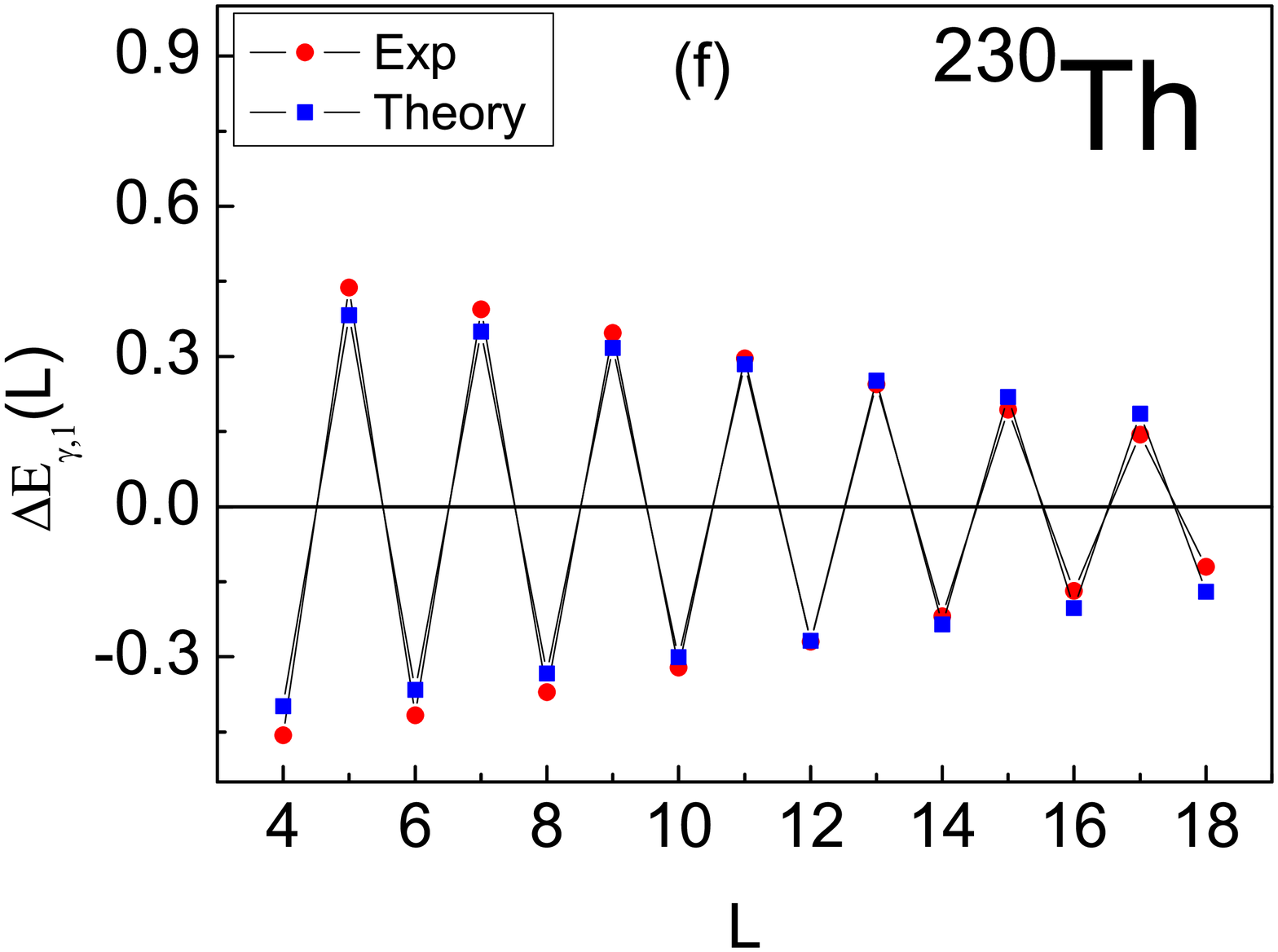}
\caption{(Color online) Theoretical and experimental staggering
function $\Delta E_{\gamma,1}(L)$ (\ref{StagL}) in $^{152}$Sm,
$^{154}$Sm, $^{148}$Nd, $^{150}$Nd, $^{226}$Ra and $^{230}$Th.}
\label{Delta}
\end{figure}

\end{widetext}

\subsection{The energy staggering}

A convenient measure for deviation from the pure rotational behavior
is the signature-splitting index $S(L)$ \cite{SI}:
\begin{equation}
S(L)=\frac{[E_{L+1}-E_{L}]-[E_{L}-E_{L-1}]}{E_{2^{+}_{1}}},
\label{SI}
\end{equation}
which vanishes for
\begin{equation}
E(L)=E_{0}+AL(L+1), \label{AL}
\end{equation}
but not for
\begin{equation}
E(L)=E_{0}+AL(L+1)+B[L(L+1)]^{2}. \label{ABL}
\end{equation}
Another quantity is also used in practice \cite{stag}
\begin{align}
\Delta E_{\gamma,1}(L)=&\frac{1}{16}(6\Delta E(L)-4\Delta E(L-1)-4\Delta E(L+1)  \notag \\
&+\Delta E(L+2)+\Delta E(L-2)), \label{StagL}
\end{align}
where $\Delta E(L)=E(L)-E(L-1)$. The staggering function
(\ref{StagL}), in contrast to (\ref{SI}), vanishes for (\ref{ABL})
and hence it represents a more sensitive measure for the deviations
of the nuclear dynamics from that of collective rotational motion.
We recall that the SU(3) limit of the spdf-IBM predicts \cite{stag}
a constant behavior for the staggering function (\ref{StagL}), thus
being unable to describe the latter.

In the present work we consider the odd-even staggering between the
states of the GSB and $K^{\pi} = 0^{-}$ band. The mapping of the
experimentally observed states of the two bands under considerations
onto the basis states of Table \ref{BS} ("stretched approximation")
establishes the relation between the quantum numbers $N$ and $L$. As
a result, the energies of the GSB can be expressed in the form
\cite{GGG}:
\begin{equation}
E(L)= \beta L(L+1)+\gamma L, \label{posen}
\end{equation}
whereas those of the $K^{\pi} = 0^{-}$ band as
\begin{equation}
E(L)= \beta L(L+1)+(\gamma + \eta)L + \xi. \label{negen}
\end{equation}
The relation between the new set of parameters entering in
Eqs.(\ref{posen}) and (\ref{negen}) and that in Eq.(15) is given in
Ref.\cite{GGG}. From the expressions (\ref{posen})-(\ref{negen}),
one can see that the energies of the GSB and $K^{\pi} = 0^{-}$ band
consist of rotational $L(L + 1)$ and vibrational $L$ terms. The
rotational interaction is with equal strength $\beta$ in both of the
bands.

The calculated and experimental staggering patterns for all
considered nuclei are illustrated in Fig.\ref{Delta}. As can be seen
the IVBM describes well the energy staggering, including the "beat
patterns" ($^{226}$Ra). The first "beat pattern" appears at the
point where the two bands are crossing. In order to be able to
describe the second "beat pattern" we assume that the states with
high angular momentum ($L \geq 20$) of the yrast band are members of
the first excited $\beta$-band. The correct reproduction of the
experimental energy staggering, including the "beat patterns", is
due to the mixing of different collective modes (see
Eqs.(\ref{posen}) and (\ref{negen})) within the framework of the
symplectic IVBM. The mixing of the two bands under consideration is
caused by the $L$-dependent interaction term $\eta L$ in
(\ref{negen}).

\subsection{Transition probabilities}

It is well known that the transition probabilities are a more
sensitive test for each model. Negative parity states of the
$K^{\pi} = 0^{-}$ band are characterized by the enhanced $E1$
transition strengths to the GSB. In the present work we consider
only the $B(E1)$ and $B(E2)$ transition probabilities concerning the
ground state and $K^{\pi} = 0^{-}$ bands.

\begin{widetext}

\begin{figure}[h]\centering
\includegraphics[width=58mm]{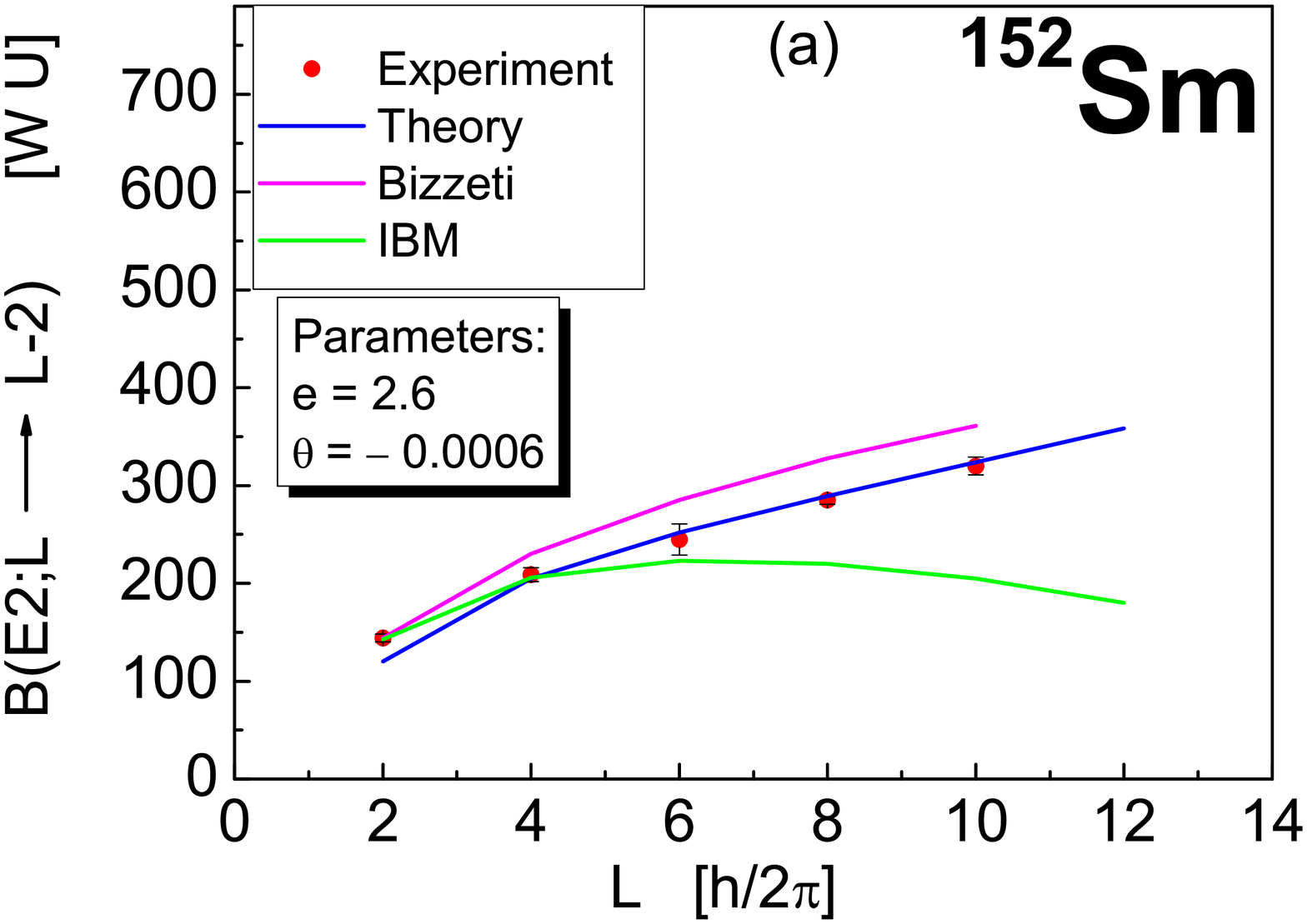}\hspace{1.mm}
\includegraphics[width=58mm]{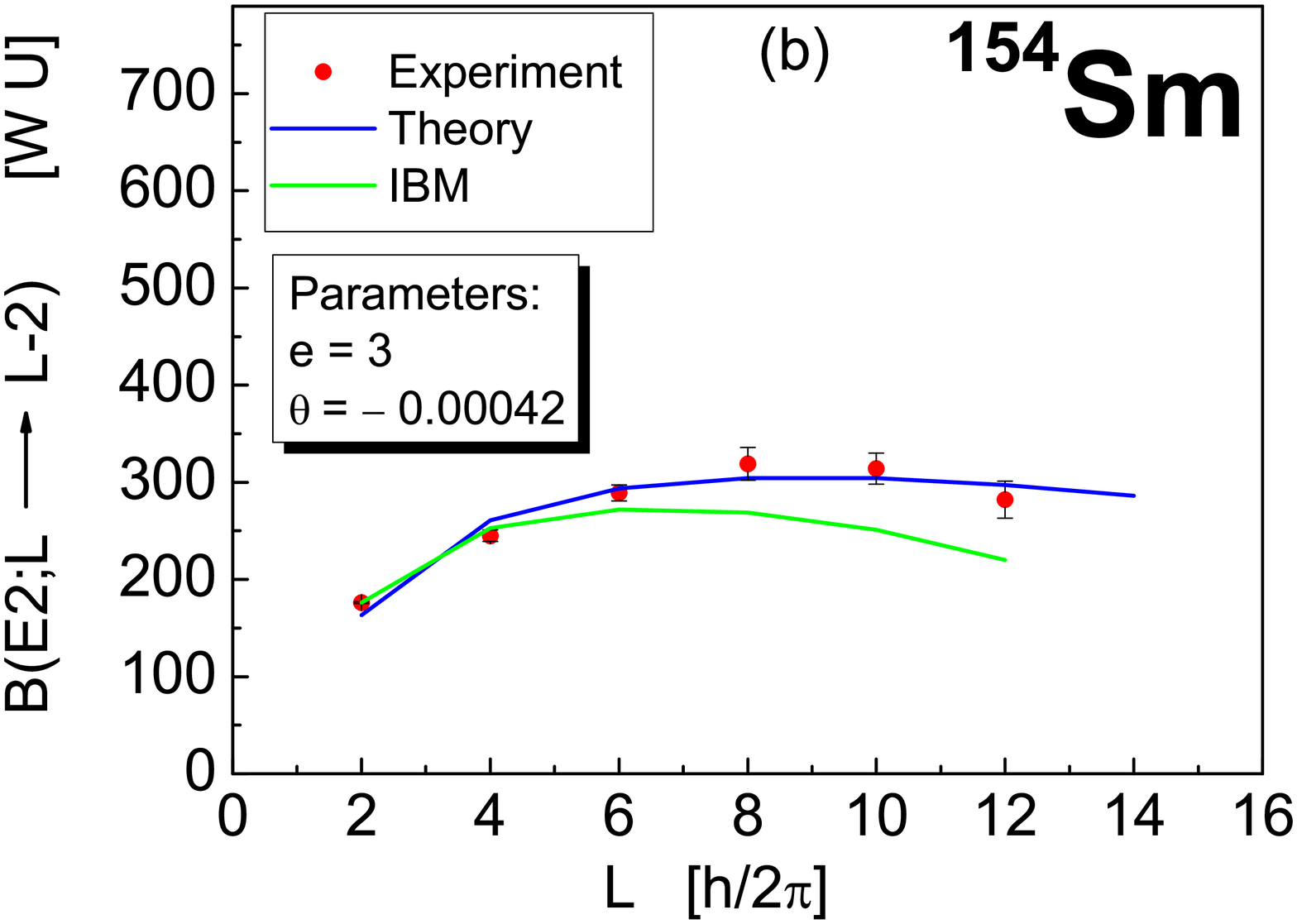}\hspace{1.mm}
\includegraphics[width=58mm]{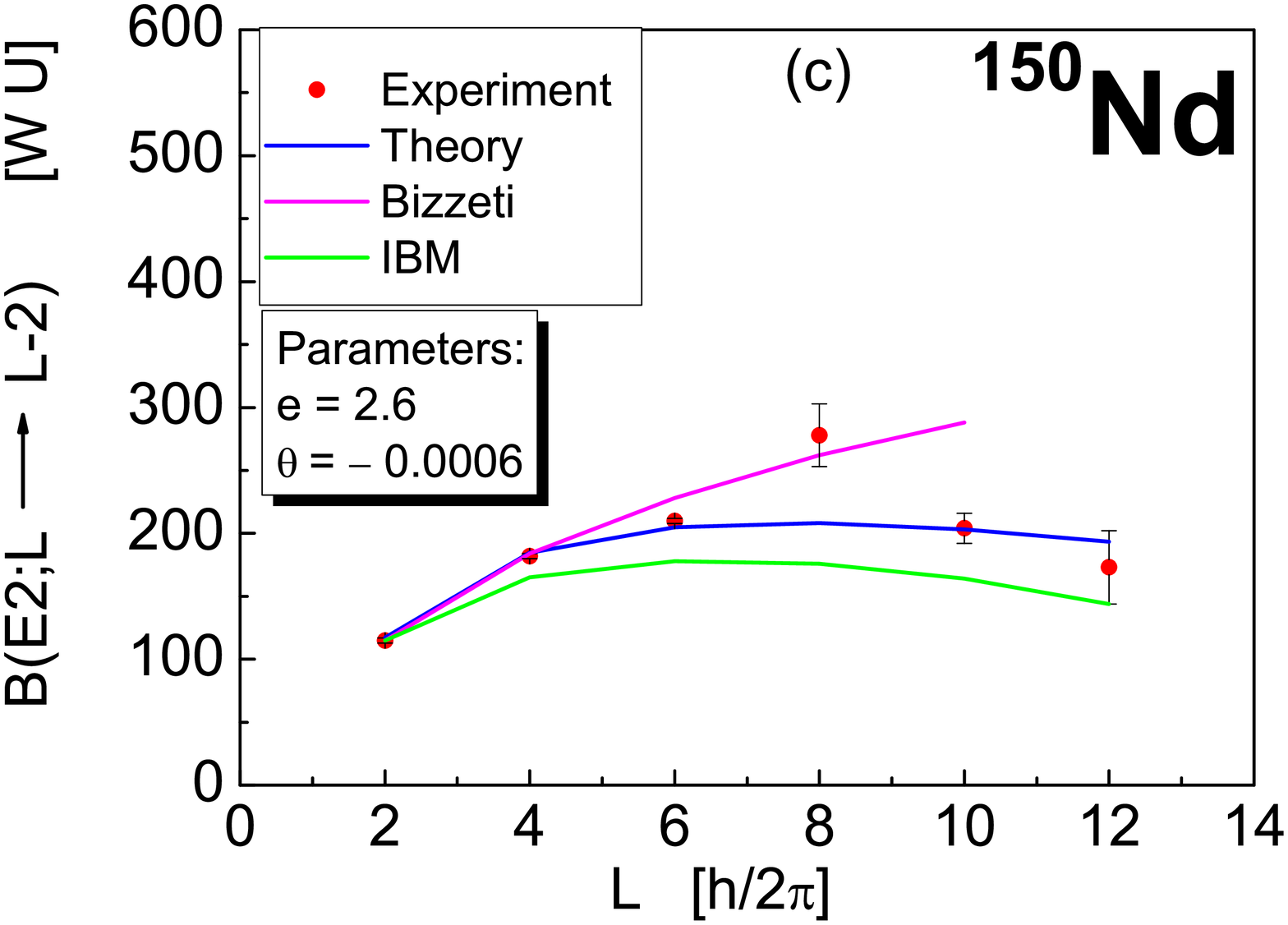}\hspace{1.mm}
\caption{(Color online) Comparison of theoretical and experimental
values for the transition probabilities of the intraband $E2$
transitions in the ground state band in $^{152}$Sm, $^{154}$Sm, and
$^{150}$Nd. For comparison, the theoretical predictions of some
other collective models are also shown.} \label{E2a}
\end{figure}

\begin{figure}[h]\centering
\includegraphics[width=70mm]{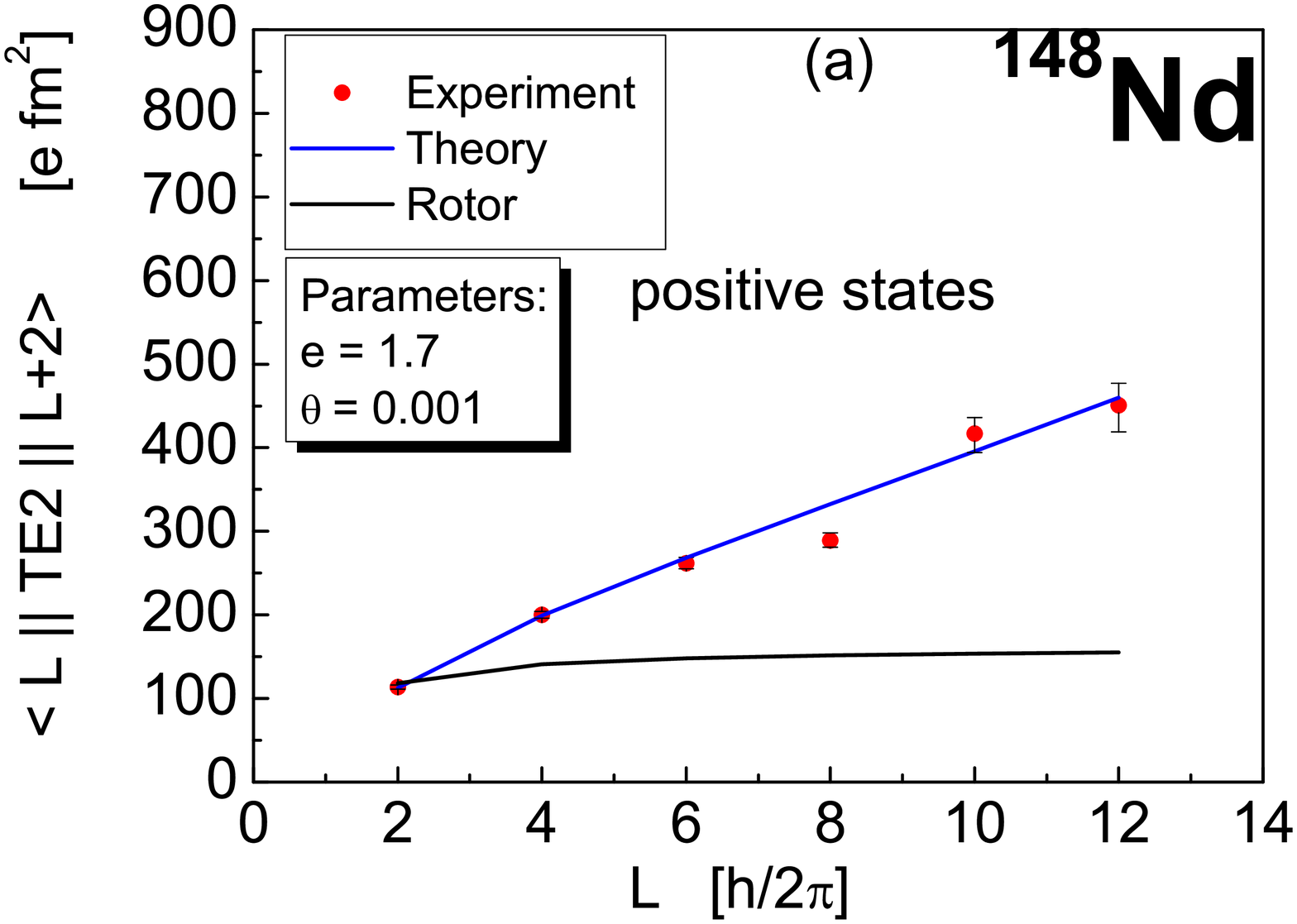}\hspace{1.mm}
\includegraphics[width=70mm]{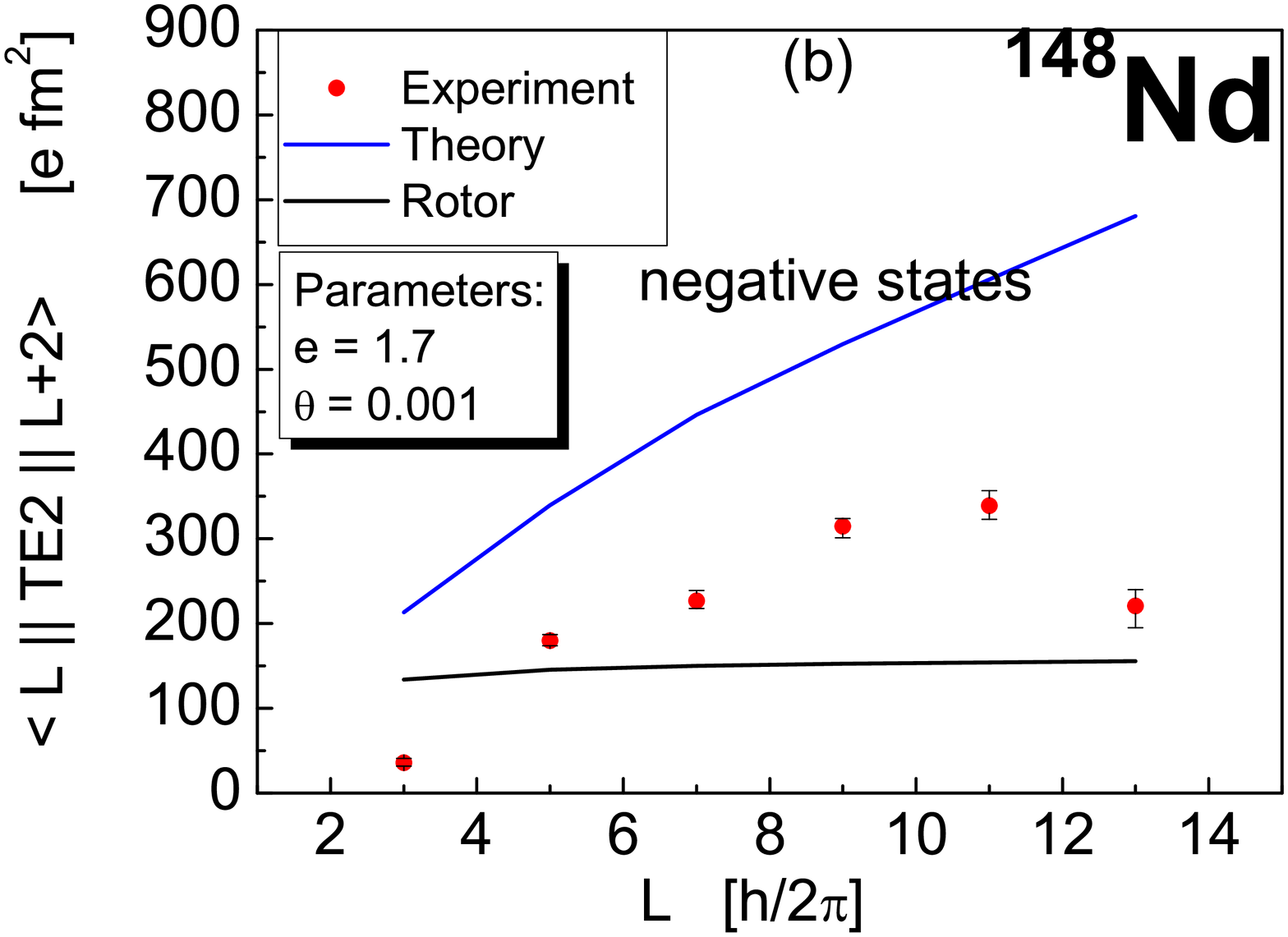}\hspace{1.mm}
\includegraphics[width=70mm]{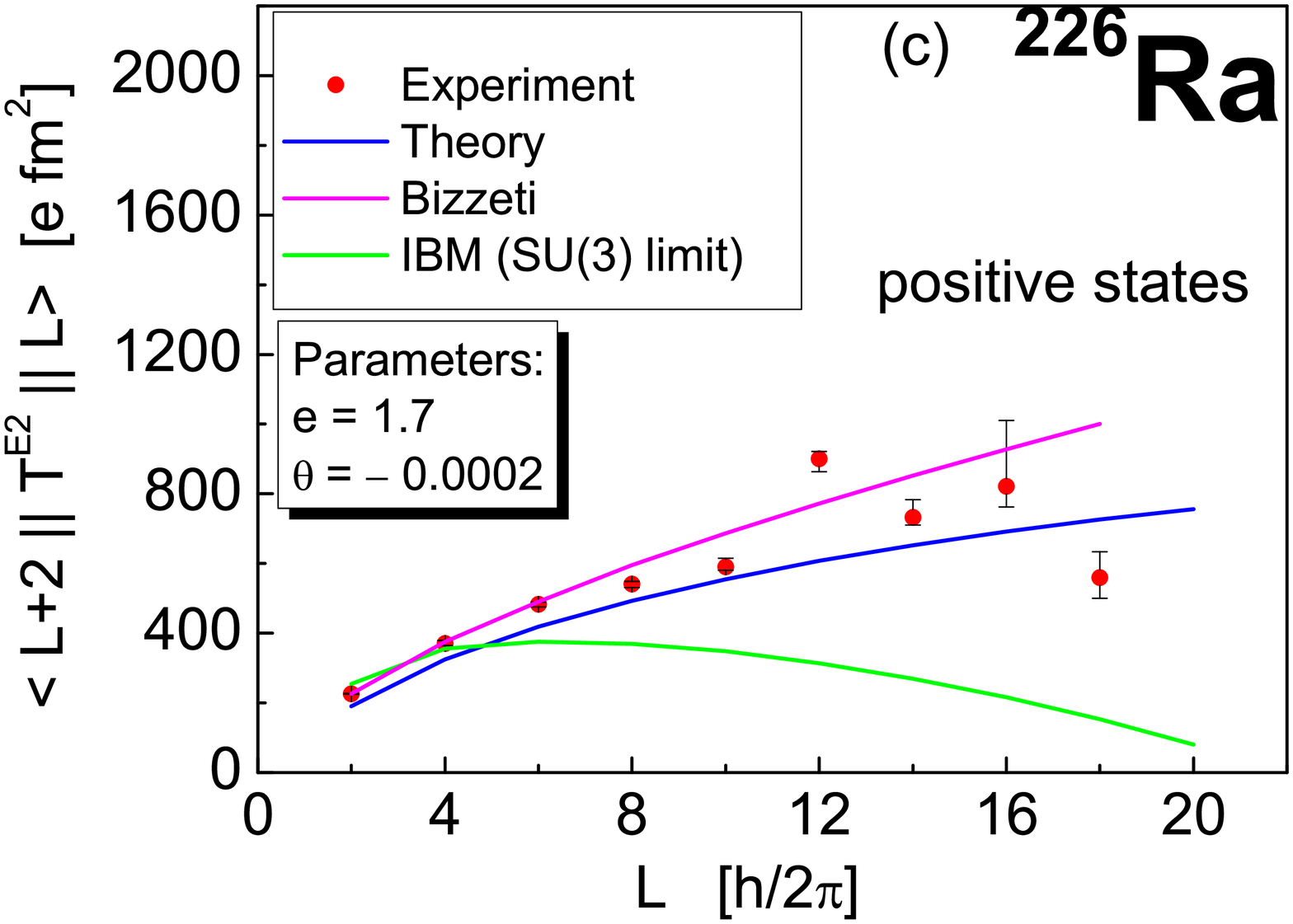}\hspace{1.mm}
\includegraphics[width=70mm]{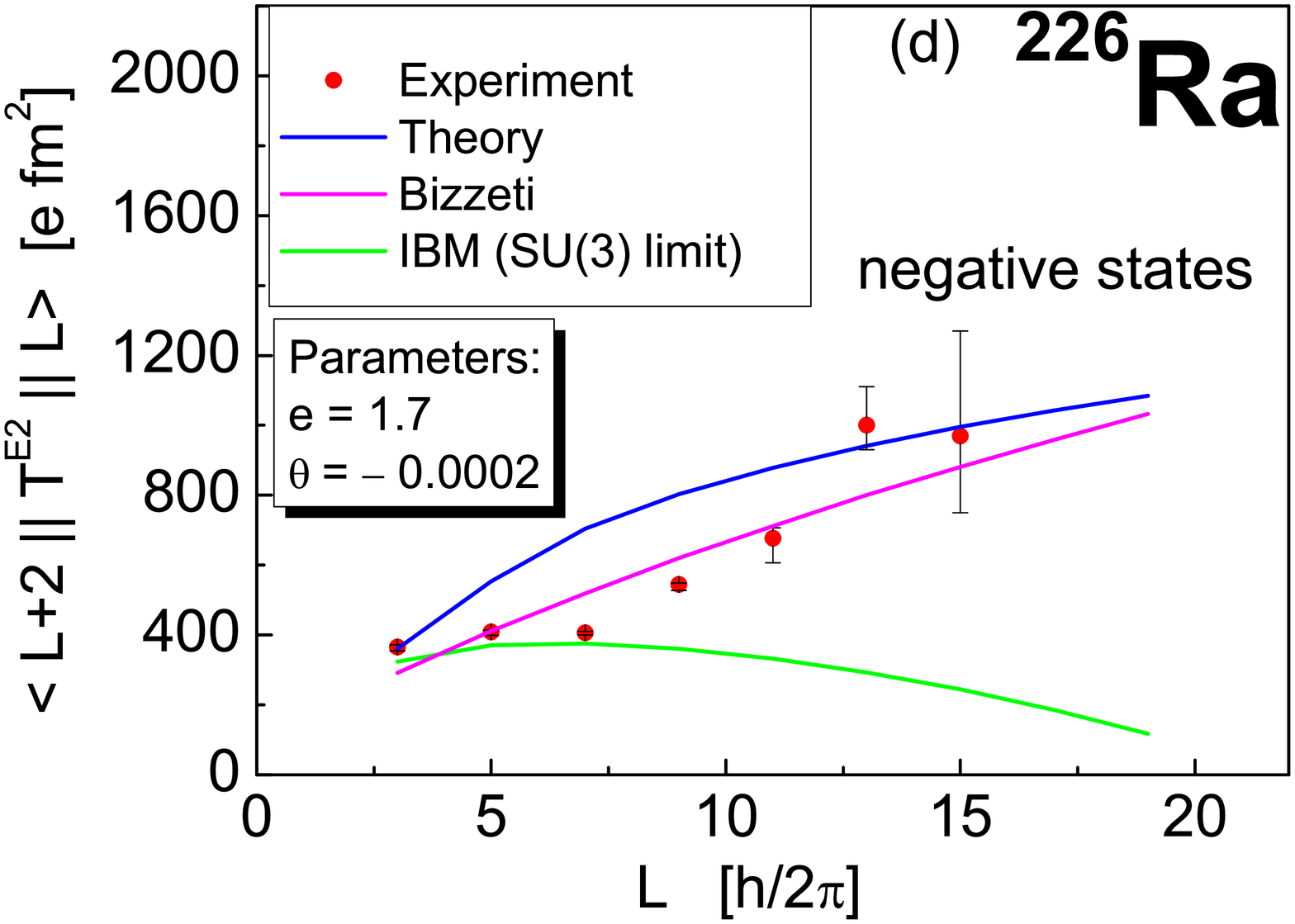}
\caption{(Color online) Comparison of theoretical and experimental
values for the matrix elements of the intraband $E2$ transitions in
the ground state band and $K^{\pi} = 0^{-}$ band in $^{148}$Nd and
$^{226}$Ra. For comparison, the theoretical predictions of some
other collective models are also shown.} \label{E2b}
\end{figure}

\begin{figure}[h]\centering
\includegraphics[width=58mm]{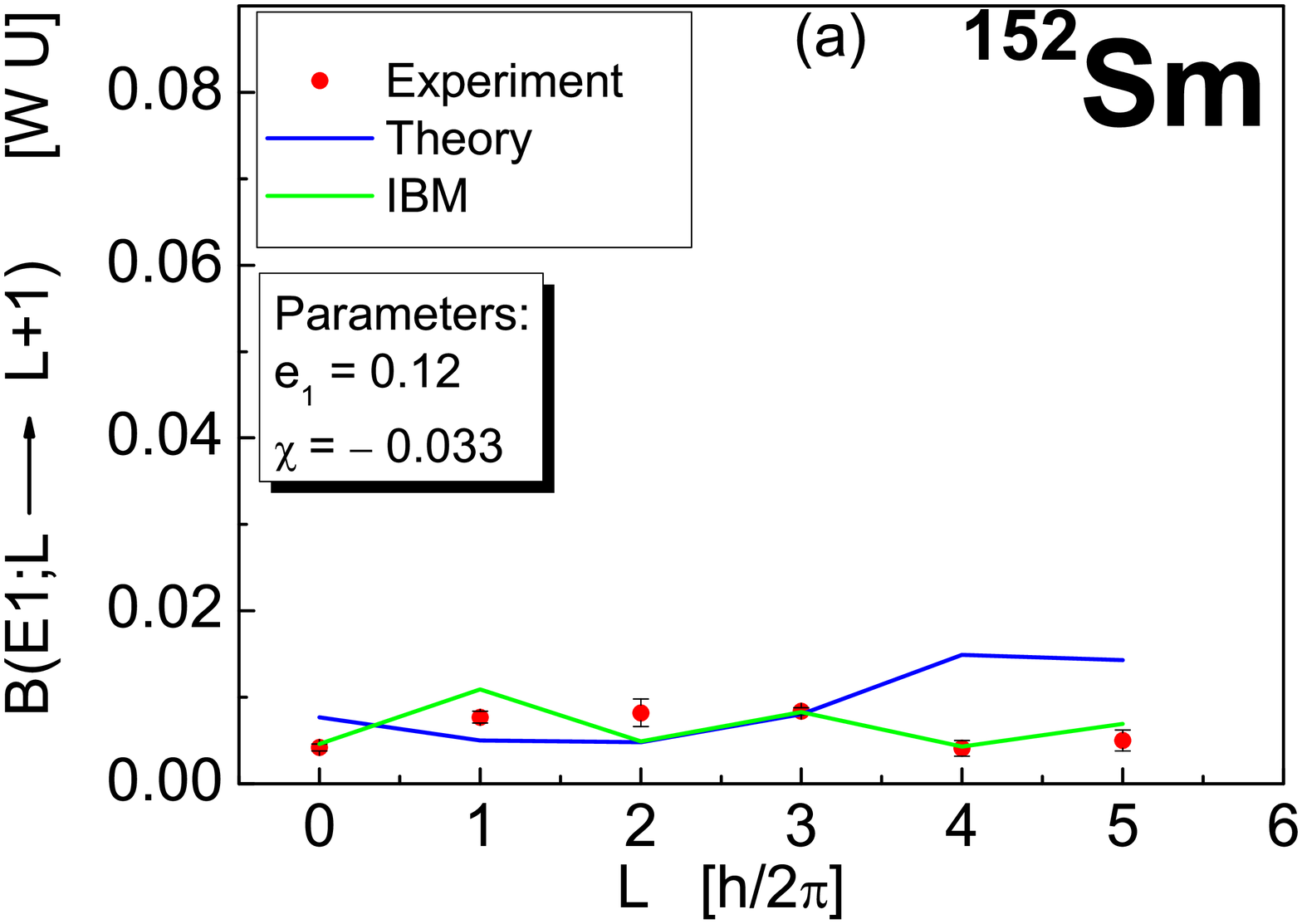}\hspace{1.mm}
\includegraphics[width=58mm]{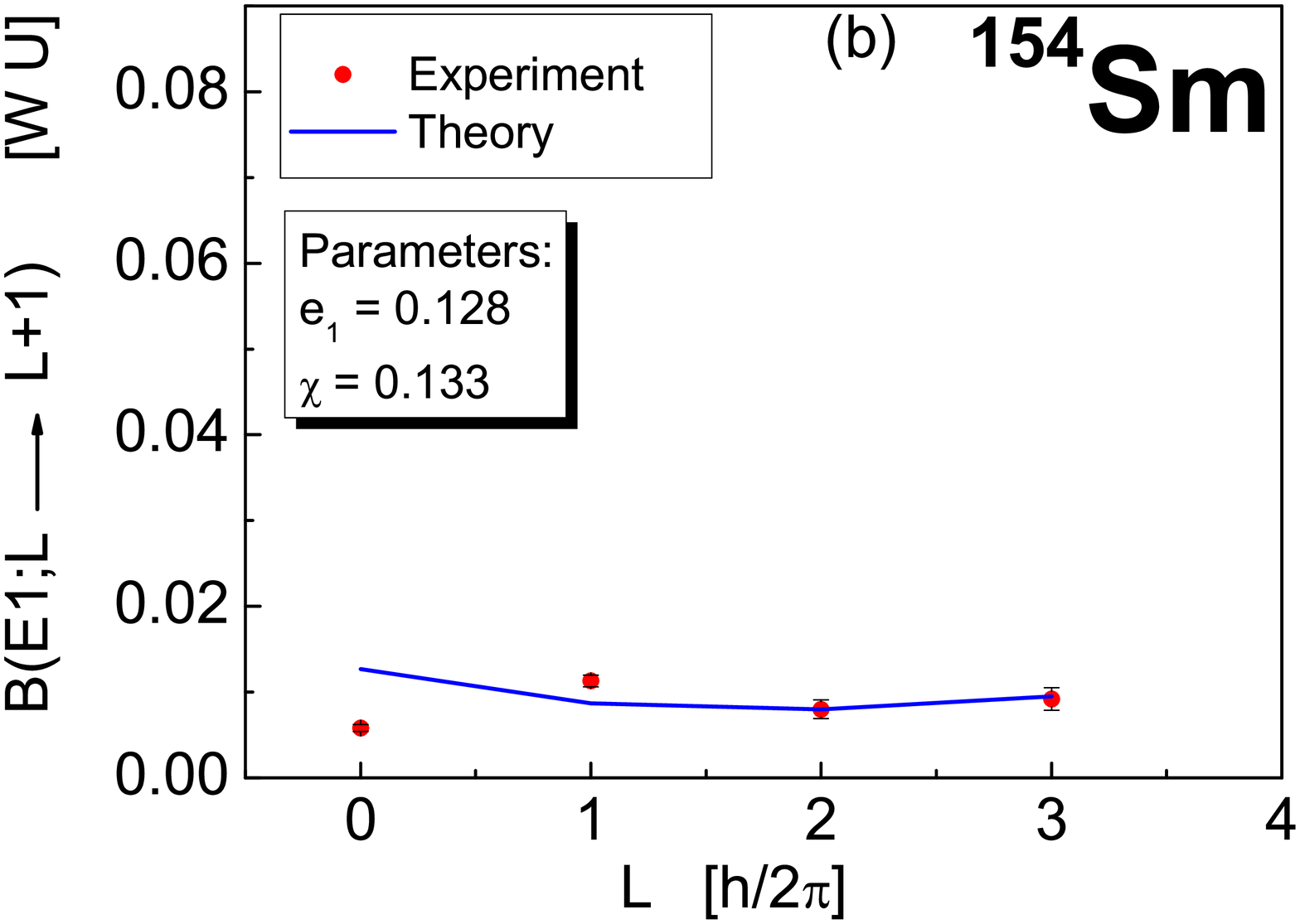}\hspace{1.mm}
\includegraphics[width=58mm]{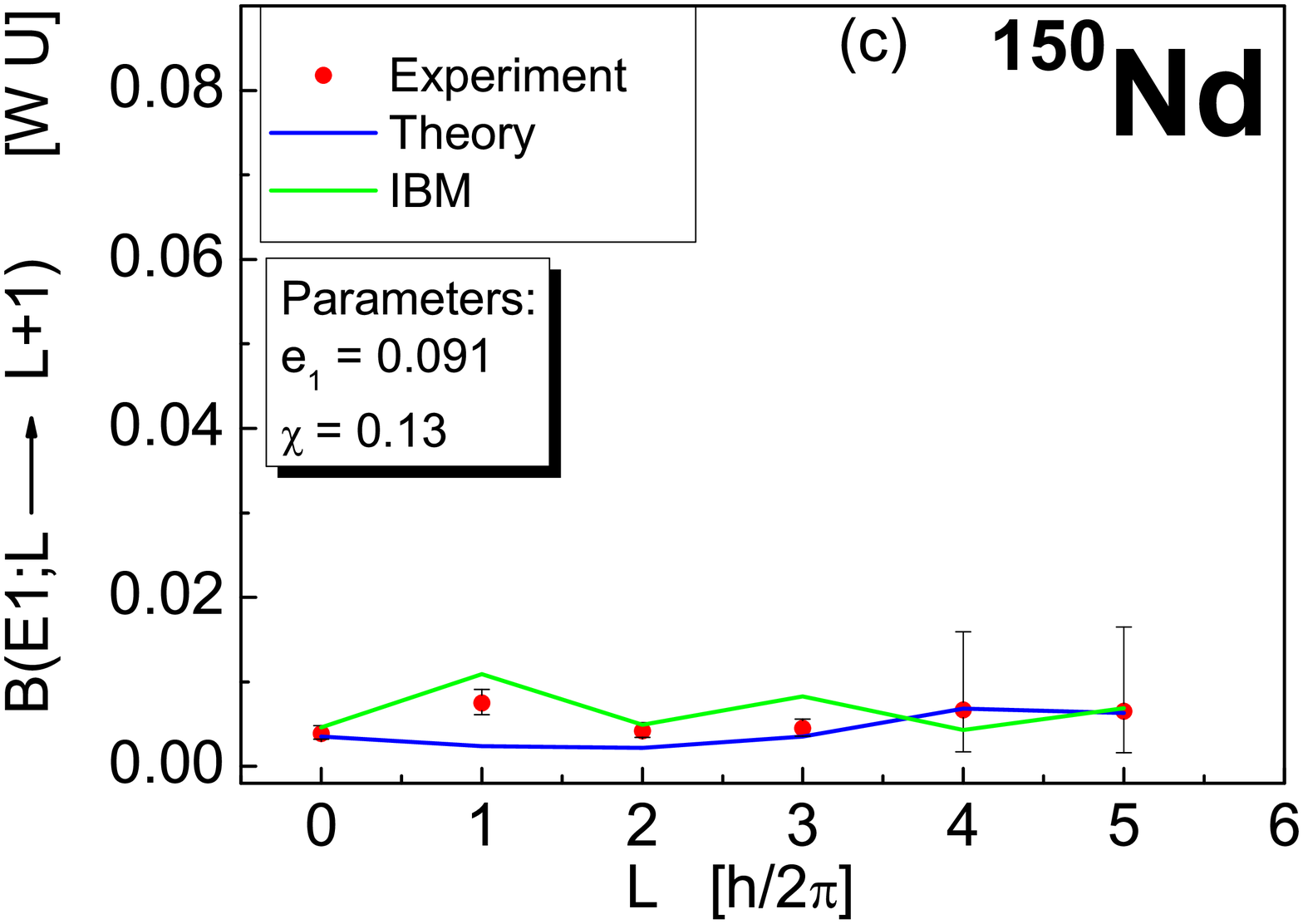}
\caption{(Color online) Comparison of theoretical and experimental
values for the transition probabilities of the interband $E1$
transitions between the states of the GSB and $K^{\pi}=0^{-}$ band
in $^{152}$Sm, $^{154}$Sm, and $^{150}$Nd. For comparison, the
theoretical predictions of some other collective models are also
shown.} \label{E1a}
\end{figure}

\begin{figure}[h]\centering
\includegraphics[width=70mm]{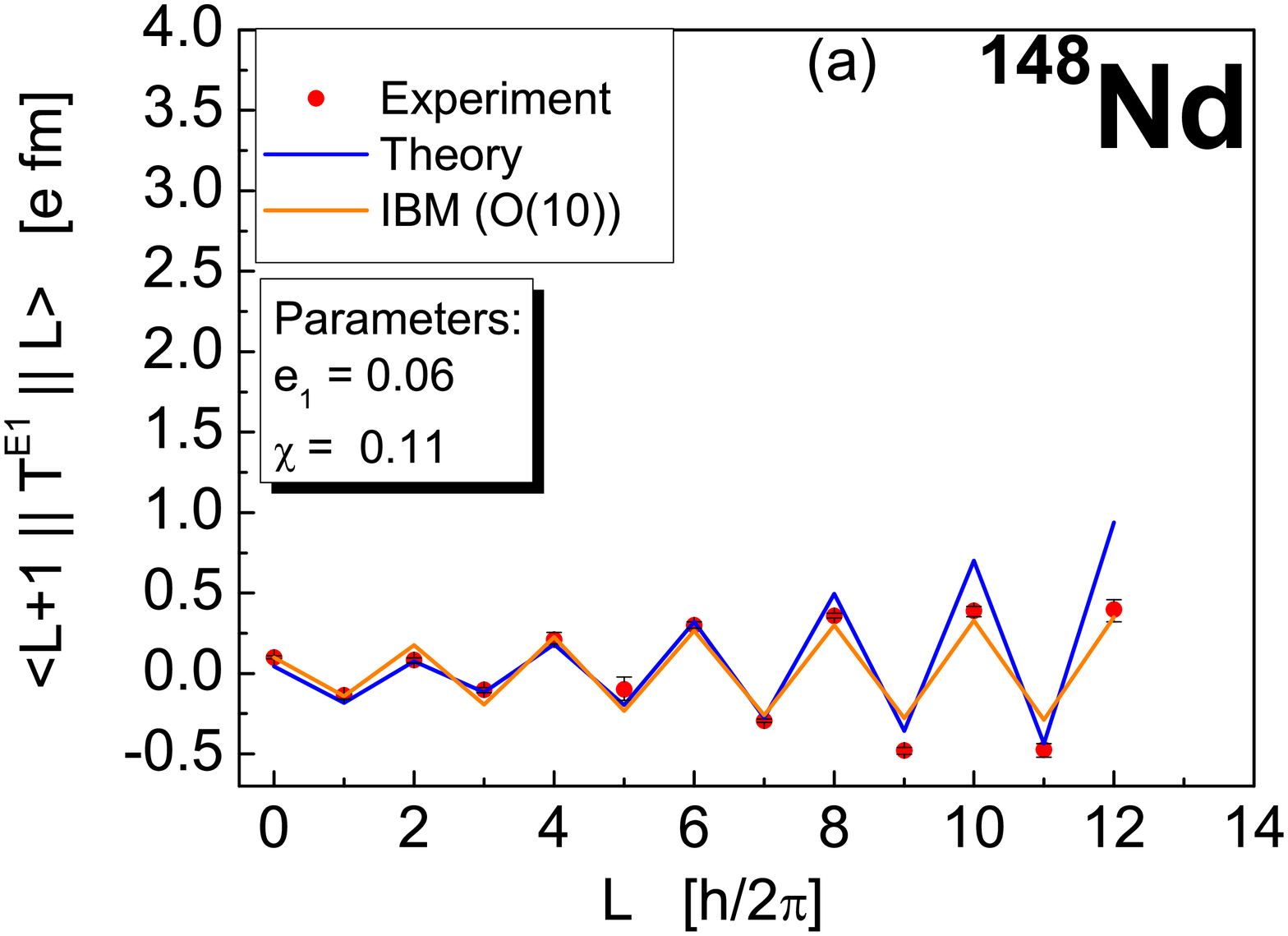}\hspace{1.mm}
\includegraphics[width=70mm]{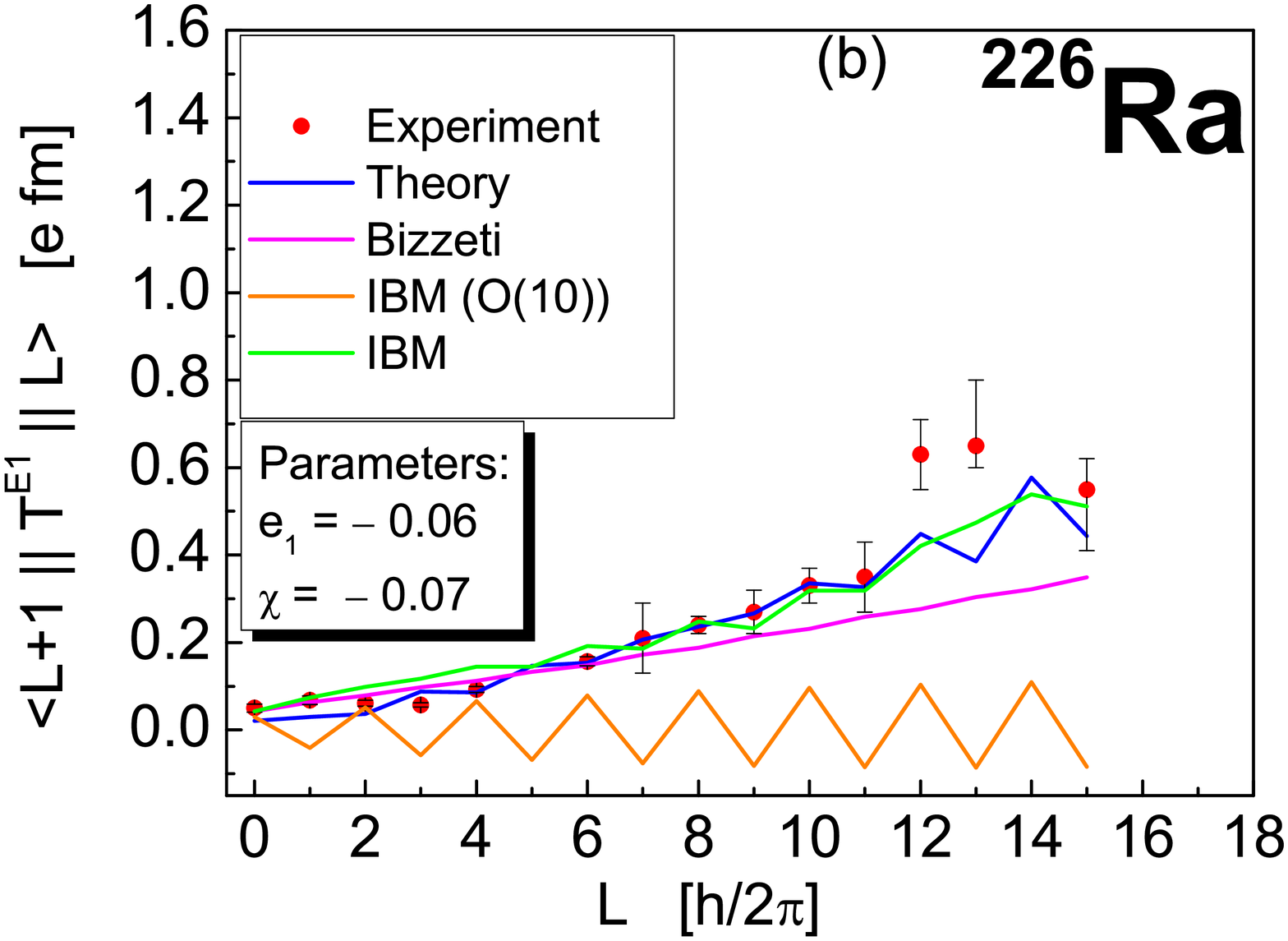}
\caption{(Color online) Comparison of theoretical and experimental
values for the matrix elements of the interband $E1$ transitions
between the states of the GSB and $K^{\pi}=0^{-}$ band in $^{148}$Nd
and $^{226}$Ra. For comparison, the theoretical predictions of some
other collective models are also shown.} \label{E1b}
\end{figure}

\end{widetext}

The transition probabilities between the collective states
attributed to the basis states of the Hamiltonian are by definition
the square of the SO(3) reduced matrix elements of the transition
operators:
\begin{equation}
B(E\lambda;L_{i}\rightarrow L_{f})=\frac{1}{2L_{i}+1}\mid \langle
\quad f\parallel T^{E\lambda}\parallel i\quad \rangle \mid ^{2}.
\label{TP}
\end{equation}
The general approach for calculating the transition probabilities
along the considered dynamical symmetry is given in Ref.\cite{TPU6},
where the $B(E2)$ transition probabilities between the states of the
GSB were calculated. Similarly, in the present work we calculate the
strengths of the intraband $E2$ transitions in both the GSB and
$K^{\pi}= 0^{-}$ band, as well as the interband $E1$ transitions
connecting the states of these two bands. In our calculations, we
use the following operators
\begin{align}
&T^{E2} = e\Big[A^{[1,-1]_{6} \ \ 20}_{(1,1)[0]_{2} \ 00}  \notag\\
&+ \theta\Big([F \times F]^{[4]_{6} \quad\quad 20}_{(0,2)[0]_{2} \
00}+[G \times G]^{[-4]_{6} \quad \ 20}_{(2,0)[0]_{2} \
00}\Big)\Big], \label{te2}
\end{align}
and
\begin{align}
&T^{E1} = e_{1}\Big[A^{[1,-1]_{6} \ \ 10}_{(1,1)[2]_{2} \ 1-1}  \notag\\
&+ \chi\Big([F \times F]^{[4]_{6} \quad\quad 10}_{(2,1)[2]_{2} \
11}+[G \times G]^{[-4]_{6} \quad \ \ \ 10}_{(1,2)[-2]_{2} \
1-1}\Big)\Big], \label{te1}
\end{align}
as transition operators for the $E2$ and $E1$ transitions,
respectively. In (\ref{te2}) and (\ref{te1}) explicit tensor
properties with respect to the reduction chain (\ref{DS}) are
written. For more details concerning the calculations we refer the
reader to Ref.\cite{TPU6}.

In Fig.\ref{E2a} we compare our theoretical results for the
transition probabilities of the intraband $E2$ transitions in the
ground state band for the three isotopes $^{152}$Sm, $^{154}$Sm, and
$^{150}$Nd. In Fig.\ref{E2b} the comparison of the theoretical
matrix elements of the intraband $E2$ transitions in both ground
state band and $K^{\pi} = 0^{-}$ band for $^{148}$Nd and $^{226}$Ra
nuclei with experiment is given. For comparison, the theoretical
predictions of some other collective models are also shown. We see
that IVBM describes reasonably well the general trend of the
experimental data. An enhancement of the theoretical $E2$ matrix
elements in the $K^{\pi} = 0^{-}$ band compared to the GSB values is
obtained. Such an enhancement was experimentally observed in
$^{144}$Ba \cite{Ba144E2}.

In Figs.\ref{E1a} and \ref{E1b} the calculated transition strengths
(matrix elements or transition probabilities) for the $E1$
transitions connecting the states of the GSB and $K^{\pi} = 0^{-}$
band are compared with experiment \cite{exp}, \cite{Ra226}
($^{226}$Ra), \cite{Nd148E1} ($^{148}$Nd) and the predictions of
some other collective models incorporating octupole or/and dipole
degrees of freedom.

An interesting zigzagging behavior of the matrix elements of the
$E1$ transitions is observed in the case of $^{148}$Nd. Such a
staggering behavior with correct phases is obtained in the framework
of the spdf-IBM if as a transition operator is used the O(10)
generator. Equivalent picture is obtained if the O(4) generator is
used as a transitional operator instead of the O(10) one. From the
Fig.\ref{E1b} one sees that IVBM is also able to describe such
staggering behavior.

\section{Conclusions}

In the present work the low-lying spectra including the first few
excited positive and negative parity bands of some heavy even-even
nuclei from the rare earth and actinide mass regions, namely
$^{152}$Sm, $^{154}$Sm, $^{148}$Nd, $^{150}$Nd, $^{226}$Ra and
$^{230}$Th, are investigated within the framework of the symplectic
Interacting Vector Boson Model with Sp(12,$R$) dynamical symmetry
group. Symplectic dynamical symmetries allow the change of the
number of excitation quanta or phonons building the collective
states providing for larger representation spaces and richer
subalgebraic structures to incorporate more complex nuclear spectra.
The theoretical predictions for the energy levels, energy staggering
and transition strengths between the collective states are compared
with experiment and some other collective models incorporating
octupole degrees of freedom. The IVBM describes well the
experimental data including some structural effects observed in the
nuclear spectra, like the "beat patterns" ($^{226}$Ra) in the energy
staggering. The results obtained for the energy levels, the energy
staggering and the transition strengths in the considered nuclei
prove the correct mapping of the basis states to the experimentally
observed ones and reveal the relevance of the used dynamical
symmetry of IVBM in the simultaneous description of the low-lying
positive and negative parity bands.

\end{document}